\documentclass[journal]{IEEEtran}
\usepackage[OT1]{fontenc} % for XeLaTeX builder
\usepackage{cite}
\usepackage{float}
\usepackage{multicol,multirow}
\usepackage{makecell,booktabs}
\usepackage{hyperref}
\usepackage{graphicx}
\usepackage{amsfonts}
\usepackage{amssymb}
\usepackage{amsmath,mathtools}
\usepackage{amsthm}
\usepackage{mathtools}
\usepackage{array}
\usepackage{color, soul}
\setstcolor{red}
\newcolumntype{P}[1]{>{\centering\arraybackslash}p{#1}}
\newcolumntype{L}[1]{>{\raggedright\arraybackslash}p{#1}}
\newcolumntype{R}[1]{>{\raggedleft\arraybackslash}p{#1}}
\usepackage{color}
\usepackage{epstopdf}
\usepackage{subcaption}
\usepackage[labelsep=period]{caption}
\usepackage{bbm}
\usepackage{bm}
\usepackage{comment}
\usepackage{forest}
\usepackage{nomencl}
\usepackage{dblfloatfix}
\usepackage{enumitem,kantlipsum}

\usepackage{etoolbox}
\usepackage[noend]{algpseudocode}
\makeatletter
\def\BState{\State\hskip-\ALG@thistlm}
\makeatother

\usepackage[ruled]{algorithm2e}
\SetKwRepeat{Do}{do}{while}

\DeclareMathOperator{\Stdev}{Stdev}

\begin{document}
\title{\huge Strategic Policymaking for Implementing Renewable Portfolio Standards: A Tri-level Optimization Approach}%    
\author{Jip~Kim,~\IEEEmembership{Student Member,~IEEE,} Sylwia Bialek, Burcin Unel
    and Yury~Dvorkin,~\IEEEmembership{Member,~IEEE
    \vspace{-2mm}}%     
}  

\maketitle
\begin{abstract}
    Appropriately designed renewable support policies can play a leading role in promoting renewable expansions and contribute to low emission goals. Meanwhile, ill-designed policies may distort electricity markets, put power utilities and generation companies on an unlevel playing field and, in turn, cause inefficiencies. This paper proposes a framework to optimize policymaking for renewable energy sources, while incorporating conflicting interests and objectives of different stakeholders. We formulate a tri-level optimization problem where each level represents a different entity: a state regulator, a power utility and a wholesale electricity market. To solve this tri-level problem, we exploit optimality conditions and develop a modification of the Column-and-Cut Generation (C\&CG) algorithm that generates cuts for bilinear terms. The case study based on the ISO New England 8-zone test system reveals different policy trade-offs that policymakers face under different decarbonization goals and implementation scenarios. 
\end{abstract}
\begin{IEEEkeywords} 
    Renewable support policy, renewable portfolio standard, state regulator, tri-level optimization, decomposition technique, column-and-cut generation
\end{IEEEkeywords}

\newcommand{\tblack}{\textcolor{black}}
\newcommand{\tblue}{\textcolor{blue}}
\newcommand{\tred}{\textcolor{red}}
% Sets
\newcommand{\setDays}{\ensuremath{\mathcal{E}}}
\newcommand{\setGens}{\ensuremath{\mathcal{I}}}
\newcommand{\setGensHat}{\ensuremath{\hat{\mathcal{I}}}}
\newcommand{\setLines}{\ensuremath{\mathcal{L}}}
\newcommand{\setNodes}{\ensuremath{\mathcal{N}}}
\newcommand{\setState}{\ensuremath{\mathcal{S}}}
\newcommand{\setTime}{\ensuremath{\mathcal{T}}}
\newcommand{\setGensRen}{\ensuremath{\mathcal{I}^{\mathrm R}}}
\newcommand{\setGensCon}{\ensuremath{\mathcal{I}^{\mathrm C}}}

\newcommand{\demFixed}{\ensuremath{{D}_{nte}}}

\section*{Nomenclature}

\subsection{Sets and Indices}
\addcontentsline{toc}{subsection}{Sets and Indices}
\begin{IEEEdescription}[\IEEEusemathlabelsep\IEEEsetlabelwidth{$n\in {\cal{N}}^{\mathrm{T/D}}a$}]
\item[$\Xi^{[\cdot]}$]{Sets of  variables, where $[\cdot]$ denotes the  regulator (R),   utility (U), wholesale market (W), master (MP), subproblem (SP) and dual wholesale market problems (DW)}
{
\item[$\Xi^{\mathrm{Aux}^{(j)}}$]{Sets of auxiliary variables, where $(j)$ is a number of the current iteration of the C\&CG algorithm}}
\item[$e\in {\cal{E}}$]{Set of representative operating days}
\item[$i\in {\cal{I}}$]{Set of existing generators}
\item[$i\in \hat{\cal{I}}$]{Set of candidate generators}
\item[${\cal{I}^{\mathrm{R}}}, {\cal{I}^{\mathrm{C}}} \subset {\cal{I}}$]{Set of renewable/controllable generators}
\item[${\cal{I}}_s \subset {\cal{I}}$]{Set of generators in state  $s$}
\item[$l\in {\cal{L}}$]{Set of transmission lines}
\item[$n\in {\cal{N}}$]{Set of transmission nodes}
\item[${\cal{N}}_s \subset {\cal{N}}$]{Set of transmission nodes in state $s$}
\item[$s\in {\cal{S}}$]{Set of states and state regulators}
\item[$t\in {\cal{T}}$]{Set of time intervals}
\item[$ {\cal{T}}^{\mathrm{on/off}}$]{Set of on- and off-peak time intervals}
\item[${r(l), o(l)}$]{Receiving/sending nodes of line $l$}
\item[${n(i)}$]{Node where generator $i$ is located}
\item[${s(i)}$]{State where generator $i$ is located}
\end{IEEEdescription}

\vspace{-3mm}
\subsection{Parameters}
\addcontentsline{toc}{subsection}{Parameters}
\begin{IEEEdescription}[\IEEEusemathlabelsep\IEEEsetlabelwidth{${G}^{\mathrm{min}/\mathrm{max}}_{it}$\!\!\!\!\!\!\!\!}]
\item[$\eta$]{Tolerance of constraint violations}
\item[$\kappa_{s}$]{Renewable portfolio standard goal of regulator $s$}
\item[$\omega_{e}$]{Probability of operating day $e$}
\item[$\rho_{ite}$]{Forecast factor of generator $i$}
\item[$\sigma_{ite}$]{Normalized standard deviation of the generation forecast error of unit  $i$ [$\mathrm{MW}$]}
\item[$\upsilon_{ite}$]{Normalized mean of the generation forecast error of unit $i$ [$\mathrm{MW}$]}
\item[$\Gamma_i$]{Minimum power output factor of controllable generator $i$, i.e. $0 \leq \Gamma_i\leq 1$ }
\item[$B^{\mathrm{P}}_s$]{Budget for renewable policies of state  $s$}
\item[$C^{\mathrm{inv}}_{i}$]{Capital cost of generator $i$ (prorated on a daily basis using the net present value approach) [$\mathrm{\$/MW}$]}
\item[$C^{\mathrm{g}}_{i}$]{Incremental cost of generator $i$ [$\mathrm{\$/MWh}$]}
\item[${D}_{nte}$]{Real power demand of node $n$ [$\mathrm{MW}$]}
\item[${F}^{\mathrm{max}}_l$]{Apparent flow limit of line $l$ [$\mathrm{MVA}$]}
\item[${G}^{\mathrm{max}\!/\!\mathrm{min}}_{it}$]{Power output limits of existing generator $i$ [$\mathrm{MW}$]}
\item[${H}^{\mathrm{max}\!/\!\mathrm{min}}_i$]{Up/down-ward ramping limits of generator $i$}
{
\item[$MCF$]{Marginal cost of public funds}
}
\item[${P}_n^{\downarrow,\mathrm{max}}$]{Apparent flow limit of the interface line into node $n$ from the transmission network [$\mathrm{MVA}$]}
\item[$X_l$]{Reactance of transmission line $l$ [$\Omega$]}
\end{IEEEdescription}

\vspace{-3mm}
\subsection{Variables}
\addcontentsline{toc}{subsection}{Variables}
\begin{IEEEdescription}[\IEEEusemathlabelsep\IEEEsetlabelwidth{$m^{(i)}_s$}]
\item[$\alpha_{it}$]{Participation factor of controllable generator $i$}
\item[$\bm{\epsilon}_{ite}$]{Forecast error of renewable generator $i$}
\item[$\theta_{nte}$]{Voltage angle of transmission node $n$}
\item[$\lambda_{nte}$]{Locational marginal price in node $n$ [$\$/\mathrm{MW}$]}
\item[$\pi_{nt}$]{Retail electricity tariff of node $n$}
\item[$ \pi_n^{\mathrm{on/off}}$]{On- and off-peak retail electricity tariff of node $n$}
\item[$\tau^{\mathrm{c}}$]{Capacity-based renewable energy incentive [$\$/\mathrm{kW}$]}
\item[$\tau^{\mathrm{e}}$]{Energy-based renewable energy incentive [$\$/\mathrm{MWh}$]}
\item[$d_{nte}$]{Flexible real power demand of node $n$ [$\mathrm{MW}$]}
\item[$f_{lte}$]{Real power flow of line $l$ [$\mathrm{MW}$]}
\item[$\bm{g}_{ite}$]{Generation output of generator $i$ [$\mathrm{MW}$]}
\item[$\overline{g}_{ite}$]{Forecast power output (offer into the wholesale market) of generator $i$ [$\mathrm{MW}$]}
\item[${g}_{ite}$]{Dispatch decisions of the wholesale market for generator $i$ [$\mathrm{MW}$]}
\item[$g^{\mathrm{max}}_{i}$]{Capacity of  generator $i$ to be built [$\mathrm{MW}$]}
\item[$p^{\downarrow}_{nte}$]{Interface power flow into node $n$ [$\mathrm{MW}$]}

\end{IEEEdescription}
{ We define random variables in \textbf{bold} and dual variables $\xi_{lts}, \underline{\gamma}_{ite}, \overline{\gamma}_{ite}, \underline{\delta}_{ite}, \overline{\delta}_{ite}$ of constraints in \eqref{LLproblem} in parentheses.}

\vspace{-3mm}
\section{Introduction}\label{Sec:intro}

The \textit{climate crisis} is the foremost priority of environmental policies around the globe, and reducing $\mathrm{CO_2}$ emissions from the electricity sector is a second-to-none curtailment instrument.
Electricity production accounts for 33\% of annual $\mathrm{CO_2}$ emissions in the U.S. and 65\% of that, i.e. 1,150 MMmt out of 1,763 MMmt, is produced by coal generators \cite{EIA_CO2}. 
Thus, many policy-makers aim   to decarbonize electricity production, relying on  a variety of policy instruments, including forced power plant retirements, carbon taxes, cap-and-trade programs, energy and capacity based incentives for carbon-free resources, renewable portfolio standards (RPS), and many others \cite{hogan2015cleaner}. A wide literature has shown that carbon pricing is the optimal tool for efficiently decreasing $\mathrm{CO_2}$ emissions from the electricity sector as well as from the whole economy \cite{aldy2010designing,fischer2008environmental,palmer2011federal}. Ideally, carbon pricing would be the main decarbonization instrument, with $\mathrm{CO_2}$ emissions priced according to the marginal damages and low-emission resources (e.g. renewables, storage, demand-side management) given necessary subsidies. For example, these subsidies can be based on the R\&D spillover rate or compensate for generation spillovers from learning-by-doing \cite{fischer2008environmental}. However, putting a price on a unit of $\mathrm{CO_2}$ emission has been politically unpopular\footnote{In the U.S., carbon pricing initiatives have failed both on the federal level (e.g. Waxman-Markey bill, Coons/Feinstein bill and many others) and state level (I-732 initiative and SB 5971 bill in Washington, SB1530 bill in Oregon, among others). In France, carbon pricing legislation was one of the factors behind the popular protest movement of the “Yellow Vests.” See also \cite{Soren2019} for an attempt to explain why carbon taxes are hard to pass.} such that in practice, only few jurisdictions maintain meaningful carbon prices \cite{Dsire}, i.e. put a price on $\mathrm{CO_2}$ emissions that comes close to the value of marginal damages from these emissions. Instead, policy-makers have widely chosen to rely on second-best policies of subsidizing carbon-free generation and investment (see \cite{hogan2015cleaner} for some  of the unintended consequences of subsidies).

Ill-designed policies can distort electricity markets and  benefit certain stakeholders over others,  even leading  to artificially prolonged economic life of certain generation types, and   result in inefficiencies of the entire market and power system. Entangled interests of stakeholders and their strategic behavior in the electricity market, along with the physical interconnection through the system, complicate the decision-making process even further. Hence,  there is a need in policy design tools, which are capable of taking into account different techno-economic perspectives of stakeholders   (e.g. regulators, power utilities, generation companies, consumers, etc.)   and physical limits on power system operations.

Among a diverse set of policy instruments for grid decarbonization \cite{hogan2015cleaner}, we narrow our attention to incentives for carbon-free resources, such as RES, given the relative prevalence and success of these incentives in the US and globally. The RES incentives fall into two categories: energy- and capacity-based. The former policies remunerate RES producers proportionally to their actual renewable production (e.g. FIT, REC), while the latter policies remunerate RES producers based on their installed capacity (e.g. PTC). Many countries use either   energy or capacity-based policies, or both of them, depending on their socio-economic circumstances \cite{murdock2019renewables}. The majority of U.S. states support RPS goals using RECs or PTCs, while Great Britain, Japan, and China have recently turned to FITs \cite{murdock2019renewables}. Regardless of the type, these policies have played  a leading role in rolling out 1,449 GW of renewables all over the world (260 GW in the U.S.) \cite{ajadi2019global}.

Methodologically, the literature on RES incentives has  been dominated by empirical data analyses, rather than model-based analyses. Huntington et al. \cite{huntington2017revisiting} compared the performance of capacity- and energy-based incentives in Spain and concluded that capacity-based incentives are more compatible with electricity markets in the considered case because they do not distort  market outcomes and generation payments, thus  not affecting the  efficiency of market competition in the considered setting. Furthermore, Newbery et al.\cite{newbery2018market} raised concerns that FITs, which are a non-competitive revenue stream that drives  RES profitability,  cause locational  market distortions affecting conventional producers. On the other hand, \"Ozdemir et al.\cite{ozdemir2019capacity} demonstrated that energy-based incentives are more cost-effective for incentive providers to achieve production goals, whereas  capacity-based incentives are more desirable in the longer term and allow for achieving  RES-rich RPS goals. Furthermore, Nicolini and Tavoni \cite{nicolini2017renewable} analyzed the effectiveness of RES incentives in five European countries and, in line with  \cite{ozdemir2019capacity}, concluded that FITs outperformed tradable green certificates (similar to RECs in the U.S.) in terms of achieving a greater RES production over the period from 2000 to 2010. Although both capacity- and energy-based RES incentives have pros and cons that have been demonstrated via empirical data analyses in \cite{huntington2017revisiting, newbery2018market, ozdemir2019capacity, nicolini2017renewable},  regulatory and legislative bodies still lack decision-support tools for (i) exogenous policymaking (i.e., optimization of incentives) and (ii) on-par comparison of different incentive types.

A critical point in designing renewable energy incentives is the need to accommodate and balance the perspectives of multiple stakeholders, as for example state regulators, power utilities, generation companies, consumers, etc. Ideally, regulators aim to select such electricity tariffs and policy incentives that maximize the social welfare in their jurisdiction, e.g. to  achieve a given RPS target in the least-cost manner, and power utilities seek to profit from supplying electricity to consumers, while ensuring  reliable electricity supply. Naturally, these objectives are conflicting because  power utilities  benefit from a higher electricity tariff, while  a higher electricity tariff leads to a greater consumer payment, which in turn reduces the social welfare. Furthermore, the wholesale electricity market typically solves a social welfare maximization or cost minimization problem, which again trade offs the  interests of power utilities and generation companies. Thus, setting renewable energy incentives without accounting for the conflicting interests of stakeholders may yield both  market distortions and inefficient renewable expansion plans. This motivates the need for developing a decision-support tool to optimally set the electricity tariff and incentives, which balances the  interests of each stakeholder.

This paper takes the perspective of the state regulator and develops a modeling framework to jointly optimize RES incentives and electricity tariffs, {under the given RPS goal and incentive budget}, while incorporating interactions between the stakeholders involved. To appropriately model this multi-stakeholder environment, the proposed framework is formulated as a tri-level (TL) optimization problem, where each level represents a different entity: a state regulator, a power utility, and a wholesale electricity market. Additionally, the level representing the  power utility internalizes flexible demands and  time-of-use (TOU) tariffs to assess the impact and value of the demand-side flexibility on achieving the RPS goals. Using this TL optimization, this paper studies a complex trade-off between the RES incentives and electricity tariffs under RPS goals and different retirement scenarios. 

In order to realistically model power system operations and, in particular, the need in balancing capacity to support the roll-out of high penetration RES levels, we model the RES stochasticity using chance constraints, as in \cite{bienstock2014chance}, which is shown to have superior computational performance over traditional scenario-based stochastic programming. However, even if chance constraints are used to alleviate some computational complexity,  commercial solvers are still not able to deal with  multi-level structure problems efficiently and, therefore, algorithmic adaptations are inevitable. Thus, to solve the proposed TL problem, we use the column-and-cut generation (C\&CG) method,  which decomposes the original problem into master and sub-problems and solves them iteratively until they reach convergence. Unlike previous C\&CG implementations in \cite{ruiz2015robust,duarte2020multi,dvorkin2017co,moreira2014adjustable}, the proposed TL problem  contains bilinear terms with the multiplication of multiple-level variables. Thus, we enhance the original C\&CG algorithm further to contain auxiliary columns for generating cuts accounting for bilinear terms.

\begin{figure}[t]
    \centering    
    \includegraphics[width=0.82\columnwidth]{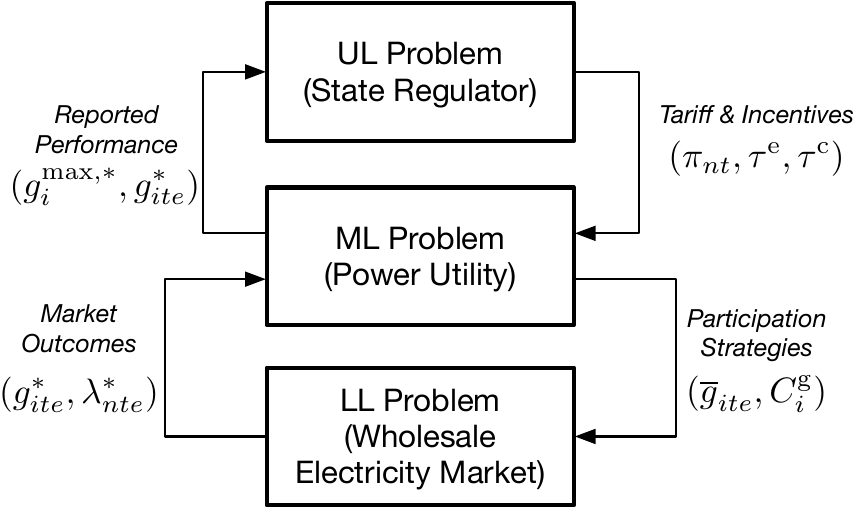}
    \caption{\small Structure of the proposed TL model. Note that $\ast$ denotes a parameterized value of decision variables from lower decision-making levels.}
    \label{fig:model}       
    \vspace{-5mm}
\end{figure} 

\vspace{-2mm}
\section{Model}\label{Sec:model}

Resorting to the TL structure, Fig.~\ref{fig:model} illustrates the proposed  model with three different stakeholders: a state regulator, a power utility and a wholesale electricity market. Each entity is represented by an optimization problem and the TL framework makes it possible to implement a leader-follower structure among entities, i.e. the state regulator is strategically more dominant than the power utility, and the power utility can act strategically in the wholesale electricity market. The goal of the upper-level (UL) problem is to represent a decision-making process of the state regulator, which sets the electricity tariffs and RES incentives and monitors performance and policy compliance of power utilities.  The middle-level (ML) problem parameterized with the UL decisions represents a decision-making process of a typical US electric power distribution utility, which decides  which generations assets, both RES and conventional, to build in order to supply electricity to its consumers and how to operate newly built and existing generation assets given representative load and renewable forecasts. The ML operational decisions include the dispatch of utility-owned assets and its participation strategy in the wholesale electricity market. Each state is assumed to have one power utility connected to the transmission network via the  power flow interface with the  limit of $P_n^{\downarrow, \max}$. Finally, the lower-level (LL) problem represents the wholesale electricity market, in which the power utility modeled in the ML problem acts strategically and the generation companies and power utilities from other states are  price takers.

\vspace{-3mm}
\subsection{UL problem (State regulator)}

In the U.S., states can set renewable policies and provide incentives ($\tau^{\mathrm e}, \tau^{\mathrm c}$) to support their implementation. At the same time, they regulate electricity tariffs ($\pi_{nt}$) so that the consumer payments are not unjustly manipulated by power utilities, while power utilities are guaranteed to recover the incurred costs to supply electricity reliably. 
Following this practice, we formulate the problem of the state regulator as:
\begin{subequations}
\begin{align}
    \begin{split}
        & \max_{\Xi^{\mathrm{R}}} {O}^{\mathrm{R}}_s\! \coloneqq\! \sum_{e\in{\cal E}} \omega_e\Big[        
        \underbrace{
            \sum_{t\in{\cal T}}\sum_{n \in {\cal N}_s}\Big(\!M_{nte} d_{nte} \!-\! \frac{1}{2} N d_{nte}^2 \!-\! \pi_{nt} d_{nte} \Big)
        }_{\text{Surplus of consumers}} \hspace{-10mm}
        \\& \hspace{5mm}
        \underbrace{
            + \sum_{t\in{\cal T}}\sum_{n\in{\cal N}_s}\pi_{nt}d_{nte}  
            + \sum_{t\in{\cal T}}\hspace{-0.5mm}\sum_{i\in\hat{\cal I}^{\mathrm R}\cup{\cal I}^{\mathrm R}}  \hspace{-3mm}\tau^{\mathrm{e}} g_{ite}           
            + \hspace{-0.5mm}\sum_{i\in\hat{\cal I}^{\mathrm R}} \tau^{\mathrm{c}} g^{\mathrm{max}}_i  
        }_{\text{Revenue of the power utility}} \hspace{-9mm}
        \\& \hspace{5mm}
        - \underbrace{
            \hspace{-1mm}\sum_{t\in{\cal T}} \Big( \hspace{-0.5mm}\sum_{n\in{\cal N}_s} \hspace{-1.5mm} \lambda_{nte} d_{nte}\hspace{-1mm}
            -  \hspace{-2mm}\sum_{i \in {\cal I}\cup\hat{\cal I}} \hspace{-1.5mm}\lambda_{n(i),t,e} {g}_{ite} \!\Big) 
        }_{\text{Cost of wholesale electricity purchased by the utility}}
        \\& \hspace{5mm}
        - \underbrace{
            \sum_{i\in\hat{\cal I}} C^{\mathrm{inv}}_i g^{\mathrm{max}}_{i}
            - \sum_{i\in{\cal I}}\sum_{t\in{\cal T}} C^{\mathrm{g}}_i g_{ite}
        }_{\text{Generation investment and operating costs}}
        \\&\hspace{5mm}         
        - \mathrm{MCF}\underbrace{ \Big(
            \sum_{t\in{\cal T}}\hspace{-0.5mm}\sum_{i\in\hat{\cal I}^{\mathrm R}\cup{\cal I}^{\mathrm R}}  \hspace{-3mm}\tau^{\mathrm{e}} g_{ite} 
            + \hspace{-0.5mm}\sum_{i\in\hat{\cal I}^{\mathrm R}} \tau^{\mathrm{c}} g^{\mathrm{max}}_i \Big)
        }_{\text{Cost of RES incentives}}\hspace{2mm}\Big]
    \end{split}\label{Eq:RegObj}\\
     & \sum_{t \in {\cal T}} \sum_{i\in{\cal I}_s^{\mathrm R}\cup\hat{\cal I}_s^{\mathrm R}} \!\!\!\!{g}_{ite} \tau^{\mathrm{e}} +   \sum_{i\in\hat{\cal I}_s^{\mathrm R}} {g}^{\mathrm{max}}_{i} \tau^{\mathrm{c}}  \le  B^{\mathrm{P}}_s,~\forall e\in{\cal E},\label{Eq:Tariff_budget}\\
     & \sum_{t \in {\cal T}} \sum_{i \in {\cal I}_s^{\mathrm R}\cup \hat{\cal I}_s^{\mathrm R}} \!\!{g}_{ite} \ge \kappa_s \sum_{t \in {\cal T}} \sum_{n\in{\cal N}_s} {d}_{nte},~\forall e\in{\cal E}, \label{Eq:StrReg_kappa}\\
     & O^{\mathrm{U}}({g}^{\max}_i, {g}_{ite}) \ge 0,\label{Eq:Radeq}\\
     & \pi_{nt} = 
     \begin{cases}
        &\pi^{\mathrm{on}}_n,\quad\forall t \in {\cal T}^{\mathrm{on}},\\
        &\pi^{\mathrm{off}}_n,\quad\forall t \in {\cal T}^{\mathrm{off}},
    \end{cases}\label{Eq:TOU_model1}
\end{align}\label{ULproblem}\end{subequations}
where $\Xi^{\mathrm{R}}=\{\pi_{nt}, \pi^{\mathrm{off}}_n, \pi^{\mathrm{on}}_n, \tau^{\mathrm{e}}, \tau^{\mathrm{c}} \ge 0\}$. 
{ Eq.~\eqref{Eq:RegObj} maximizes the expected social welfare over representative days $e$ with probability $\omega_e$. This welfare function encompasses the surplus of consumers, revenue of the power utility, cost of providing electricity (generation investment and operating cost, wholesale costs of imports), and cost of RES incentives provided to support renewable expansion.
In the following, we assume that the state regulator can raise money for RES incentives without creating welfare losses, i.e. the marginal cost of public funds (MCF) equals one, as common in economics literature. This assumption is true for many settings  \cite{Jacobs2018,BENTO2018}, however, it can be adjusted if actual costs of raising public funds have a non-unitary MCF, (e.g. MCF = 0.9, 1.2, etc) \cite{Barrios2013}.
}
The capacity-based incentive ($\tau^{\mathrm c}$) has a net present value  prorated on a daily basis. The renewable policy budget limit is imposed in Eq.~\eqref{Eq:Tariff_budget}, and is procured externally from the optimization and is decoupled from the electricity tariff ($\pi_{nt}$). The RPS target constraint, representing a given renewable policy, is in Eq.~\eqref{Eq:StrReg_kappa} and the revenue adequacy (non-negative profit) constraint for the power utility is imposed in Eq.~\eqref{Eq:Radeq} to ensure the power utility is not exposed to the deficit by investing in RES, where $O^{\mathrm{U}}$ is the objective function of the utility defined in Eq.~\eqref{Utility_Obj}. The non-negativity constraint on profits of the utility does not preclude retirement of the assets -- as we do not model decommission costs or annual, fixed maintenance costs, retirement of a power plant is equivalent to having zero capacity utilization rate of that power plant. The electricity tariff is modeled as a time-of-use rate in Eq.~\eqref{Eq:TOU_model1} while ${\cal T}^{\mathrm{on/off}}$ represent on- and off-peak time sets.

{ If the state regulator is unable to identify  some components of the social welfare objective function given in \eqref{ULproblem}, for instance because  electricity demand functions are unknown,  alternative objective constructs can be used. For example,  in our setting, a total payment minimization can be adopted as a proxy. This proxy includes only components, which are under control of the regulator and, therefore, invokes less assumptions.  We discuss the similarity between these two objectives and compare their performance in Appendix \ref{Appendix_proxy}.}

\vspace{-2mm}
\subsection{ML problem (Power utility)}
The ML problem models the power utility.  First, we consider  flexible demand and  generation models.   

\subsubsection{Demand} Demand elasticity  can be implicitly modeled in the ML optimization by deriving optimality conditions of the surplus maximization problem of aggregated consumers. An aggregated utility function of electricity consumers is  non-decreasing and concave and modeled as \cite{samadi2010optimal}:
\begin{subequations} \begin{align}
    & U(d_{nte}) \!= \!
        \begin{cases}
            M_{nte} d_{nte} - \frac{N}{2} d_{nte}^2 \hspace{2mm} \text{if} \quad 0\le d_{nte} \le \frac{M_{nte}}{N},\\
            \frac{M_{nte}^2}{2N} \hspace{22.2mm} \text{if} \quad  d_{nte} \ge \frac{M_{nte}}{N},
        \end{cases}\hspace{-4mm}
\end{align}\end{subequations}
where $M$ and $N$ are pre-determined parameters. 
Given electricity tariff $\pi_{nt}$, the maximized consumer surplus is:
\begin{subequations} \label{eq:flexible_demand}\begin{align}
    & \max\hspace{1mm} \text{CS}_{nte} \coloneqq(M_{nte} d_{nte} - \frac{1}{2} N d_{nte}^2) - \pi_{nt} d_{nte}
\end{align}
Then, we derive a closed-form demand  as a function of the electricity tariff ($d_{nte}(\pi_{nt})$) via the first-order optimality condition and by assuming non-zero marginal utility of electricity (setting upper bound $\overline{d}_{nte}=M_{nte}/N$):
\begin{align} 
    &\frac{\partial {\text{CS}_{nte}}}{\partial d_{nte}} = (M_{nte} - \pi_{nt}) -N d_{nte} = 0,
    \\&d_{nte}(\pi_{nt}) = \frac{M_{nte} - \pi_{nt} }{N}.  \label{eq:demand_simple}
\end{align}\end{subequations}
Given \eqref{eq:demand_simple},  the demand function depends on parameters $M$ and $N$, which are external to the optimization problem, and the value of tariff chosen by the UL optimization. If the power utility implements a demand-response program with incentives for consumers to change their power consumption, the resulting demand in \eqref{eq:demand_simple}  can be further adjusted. For instance, if $\Delta \pi_{nt} \geq 0$ denotes the demand-response incentive, the resulting demand in \eqref{eq:demand_simple}  will be set to $d_{nte}(\pi_{nt} + \Delta \pi_{nt}) = \frac{M_{nte} - \pi_{nt} - \Delta \pi_{nt} }{N}.$

\subsubsection{Generation}
The uncertain RES generation output is modeled through forecast error variables $\bm{\epsilon}_{ite}$ with Gaussian distribution \footnote{The Gaussian distribution is used in this work due to its mathematical tractability, while other probability distributions  can be adopted to model forecast errors with more accuracy (e.g. Student's t, Logistic, Cauchy-Lorentz distributions, { see Appendix. \ref{Appendix_NonGuassian}).}}
${\cal G}({\mathbb{E}(\bm{\epsilon}_{ite})}, {\mathrm{Stdev}(\bm{\epsilon}_{ite})}^2)$ as:
\begin{subequations}\label{Eq:Random_model}\begin{align}
\bm{\epsilon}_{ite} \!\!\sim\!\! \begin{dcases}
    &\hspace{-3mm} {\cal G}({G^{\mathrm{max}}_i}\upsilon_{ite}, ({G^{\mathrm{max}}_i} \sigma_{ite})^2) ,\hspace{0.0mm}\forall i \in{\cal I}^{\mathrm R}_s, t \in {\cal T}\!, e\in{\cal E},\hspace{-3mm}
    \\&\hspace{-3mm} {\cal G}({g^{\mathrm{max}}_i}\upsilon_{ite}, ({g^{\mathrm{max}}_i} \sigma_{ite})^2),\hspace{2mm}\forall i \in\hat{\cal I}^{\mathrm R}_s, t \in {\cal T}\!, e\in{\cal E},\hspace{-3mm}
\end{dcases}
\end{align}
where $\upsilon_{ite}$ and $\sigma_{ite}$ are the normalized mean and standard deviation of the forecast error, and ${G^{\mathrm{max}}_i}$ and ${g^{\mathrm{max}}_i}$ are the generation capacity of existing (${\cal I}^{\mathrm R}_s$) and candidate ($\hat{\cal I}^{\mathrm R}_s$) RES generators. 
Next, the forecast power output of RES generators, $\overline{g}_{ite}$, can be modeled with a forecast factor $\rho_{ite}\in [0,1]$ as:
\begin{align}
\overline{g}_{ite} \!=\! \begin{dcases}
    &\hspace{-3mm} \rho_{ite} G_i^{\mathrm{max}} ,\hspace{5mm}\forall i \in{\cal I}^{\mathrm R}_s,t \in {\cal T}, e\in{\cal E}, 
    \\&\hspace{-3mm} \rho_{ite} g_i^{\mathrm{max}},\hspace{6mm}\forall i \in\hat{\cal I}^{\mathrm R}_s, t \in {\cal T}, e\in{\cal E},
\end{dcases}
\end{align}
Then, the output of RES generators follows as:
\begin{align}    
    & \bm{g}_{ite} = \overline{g}_{ite} + \bm{\epsilon}_{ite}, \quad\forall i \in {\cal I}^{\mathrm{R}}_s\cup\hat{\cal I}^{\mathrm{R}}_s, t \in {\cal T}, e\in{\cal E}, \label{Eq:uncertainty_model_n}\vspace{-3mm}
\end{align}
Since forecast errors result in a real-time mismatch between the power produced and consumed, controllable generation resources must offset these deviations. In practice, the real-time affine control is used, \cite{jaleeli1992understanding}, and modeled as:
\begin{align}
    & \bm{g}_{ite} = \overline{g}_{ite} \!-\! \alpha_{it} \!\!\!\!\!\!\!\!\!\!\sum_{j \in {\cal I}^{\mathrm R}_{s(i)}\cup\hat{\cal I}^{\mathrm R}_{s(i)}} \!\!\!\!\!\!\!\!\!\!\bm{\epsilon}_{jte},~\forall i \in {\cal I}^{\mathrm C}_s\!\cup\!\hat{\cal I}^{\mathrm C}_s, t \in {\cal T}, e\in{\cal E}, \label{Eq:affine_control}
\end{align}%
\end{subequations}%
where $\sum_{i\in{\cal I}^{\mathrm C}_s\cup\hat{\cal I}^{\mathrm C}_s} \alpha_{it} \!=\! 1$ ensures the sufficiency of procured balancing resources. Synchronized generators continuously offset the power mismatch based on  parameter $\alpha_{it}$, which regulates the participation of each generator in balancing. The value of $\alpha_{it}$ can be     set ahead of time, e.g. $\alpha_{it}=1/{\text{card}({\cal{I}}^{\mathrm{C}}_s\cup \hat{\cal{I}}^{\mathrm{C}}_s})$, or optimized as in \cite{bienstock2014chance}.

\subsubsection{Power utility model}
We assume the power utility company optimizes its generation expansion plan and optimizes its operations, including the participation strategy in the wholesale electricity market as follows:
\begin{subequations}\begin{align}
    \begin{split}             
        &\max_{\Xi^{\mathrm{U}}} O^{\mathrm{U}}\!\coloneqq\!
        \hspace{-0.0mm}\sum_{e\in{\cal E}} \omega_e \Big[\sum_{t\in{\cal T}}\sum_{n\in{\cal N}_s}\pi_{nt}d_{nte}  
        + \sum_{t\in{\cal T}}\hspace{-0.5mm}\sum_{i\in\hat{\cal I}^{\mathrm R}\cup{\cal I}^{\mathrm R}}  \hspace{-3mm}\tau^{\mathrm{e}} g_{ite}   \hspace{-5mm}
        \\&\hspace{0mm}
        + \hspace{-0.5mm}\sum_{i\in\hat{\cal I}^{\mathrm R}} \tau^{\mathrm{c}} g^{\mathrm{max}}_i  
        - \hspace{-1mm}\sum_{t\in{\cal T}} \Big( \hspace{-0.5mm}\sum_{n\in{\cal N}_s} \hspace{-1.5mm} \lambda_{nte} d_{nte}\hspace{-1mm}
        -  \hspace{-2mm}\sum_{i \in {\cal I}\cup\hat{\cal I}} \hspace{-1.5mm}\lambda_{n(i),t,e} {g}_{ite} \!\Big) \hspace{-5mm}
        \\&\hspace{0mm}
        - \sum_{i\in\hat{\cal I}} C^{\mathrm{inv}}_i g^{\mathrm{max}}_{i}
        - \sum_{i\in{\cal I}}\sum_{t\in{\cal T}} C^{\mathrm{g}}_i g_{ite}\Big]\hspace{-5mm}
        \label{Utility_Obj}
    \end{split}\\
    & \mathbb{P}[\bm{g}_{ite} \le {G}_{i}^{\mathrm{max}}]\ge1-\eta,\quad\forall i \in {\cal I}^{\mathrm{C}}_s, t \in {\cal T}, e\in{\cal E},\label{Eq:gen_UB}\\
    & \mathbb{P}[\bm{g}_{ite} \ge {G}_{i}^{\mathrm{min}}]\ge1-\eta,\quad\forall i \in {\cal I}^{\mathrm{C}}_s, t \in {\cal T}, e\in{\cal E}, \label{Eq:gen_LB}\\
    & \mathbb{P} [\bm{g}_{ite} \le {g}_{i}^{\mathrm{max}}] \ge 1 - \eta,\quad\forall i \in {\cal \hat{I}}^{\mathrm{C}}_s, t \in {\cal T}, e\in{\cal E},  \label{Eq:gen_UB_new}\\
    & \mathbb{P} [ \bm{g}_{ite} \ge \Gamma_i {g}_{i}^{\mathrm{max}}] \ge 1 - \eta,\quad\forall i \in {\cal \hat{I}}^{\mathrm{C}}_s, t \in {\cal T}, e\in{\cal E},  \label{Eq:gen_LB_new}\\
    & {H}^{\mathrm{min}}_{i} \!\!\le\! \overline{g}_{ite}\!-\!\overline{g}_{i,t-1,e} \! \le \!{H}^{\mathrm{max}}_{i}\!\!,~\!\forall i \!\in \! {\cal I}^{\mathrm{C}}_s\!\cup\!\hat{\cal I}^{\mathrm{C}}_s, t \in \!{\cal T}\!, e\in\!{\cal E},\hspace{-3mm}\label{Eq:ramp}\\      
    & \sum_{i \in {\cal I}_s \cup {\hat{\cal{I}}_s}} \!\!\!\overline{g}_{ite} + \!{p}_{nte}^{\downarrow}  = \!d_{nte} ,\quad\forall t \in {\cal T},{n\in{\cal N}_s}, e\in{\cal E}, \label{TSO_P_balance2}\\
    & -\!{P}_n^{\downarrow,\mathrm{max}} \!\!\le {p}^{\downarrow}_{nte} \le {P}_n^{\downarrow,\mathrm{max}},~ \forall t \in {\cal T}\!, n\in{\cal N}_s, e\in{\cal E},\label{Eq:pdown_ULB}    
\end{align}\label{MLproblem}\end{subequations}
where ${\Xi^{\mathrm{U}}=\{g^{\mathrm{max}}_i, \overline{g}_{ite} \ge 0\}}$. Objective function \eqref{Utility_Obj} maximizes the  profit of the power utility over the operating horizon $t\in{\cal T}$ for each representative day $e$, which includes the payment collected from consumers, revenue from the wholesale market participation and renewable energy incentives provided by the state regulator. Investment cost $C^{\mathrm{inv}}_i$ and capacity-based incentive $\tau^{\mathrm c}$ are prorated on a daily basis using the net present value approach as in \cite{6936940}. Eqs.~\eqref{Eq:gen_UB}-\eqref{Eq:gen_LB_new} limit power outputs of existing and candidate resources, while the RES uncertainty is captured via chance constraints with random variable $\bm{g}_{ite}$. Eq.~\eqref{Eq:ramp} limits the up- and downward ramping rates for controllable generators. Finally, Eq.~\eqref{TSO_P_balance2} enforces the supply demand balance, while Eq.~\eqref{Eq:pdown_ULB} enforces the power flow limit ($p^{\downarrow}_{nte}$) on the power flow between the power utility and transmission network (and  the  wholesale electricity market). Note that network constraints among nodes $n\in{\cal N}_s$ within each state $s$ are not imposed in Eq.~\eqref{MLproblem}, and the power flows are constrained in the wholesale market problem.

\vspace{-2mm}
\subsection{LL problem (Wholesale market)}
The wholesale electricity market for each representative day $e\in{\cal E}$, which includes the strategic participant  modeled in the ML problem and other non-strategic participants, solves:
\begin{subequations}\begin{align}
    \begin{split}            
        &\max_{\Xi^{\mathrm{W}}} O^{\mathrm{W}}_{e}\!\coloneqq\!
       \sum_{t\in{\cal T}} \sum_{i \in {\cal I}\cup\hat{\cal I}} - C_i^{\mathrm{g}} {g}_{ite}\label{WM_Obj}
    \end{split}\\          
    &(\xi_{lte}):\!\!\!\quad f_{lte} = \frac{1}{X_l} (\theta_{o(l),te}-\theta_{r(l),te}),~\forall {l \in {\cal L}},~t\in{\cal T},\label{TSO_DSPF}\\    
    \begin{split}
        &(\lambda_{nte}):\hspace{-1mm}\sum_{i \in {\cal I}_n\cup\hat{\cal I}_n } \hspace{-1mm}{g}_{ite}  + \hspace{-1mm}\sum_{l \vert r(l) = n } \hspace{-1mm} f_{lte} \!-\hspace{-1mm}\sum_{l \vert o(l) = n } \hspace{-1mm} f_{lte} \!
        \\&\hspace{26mm}=
        \begin{dcases}
            &\hspace{-3mm}d_{nte},\quad\forall n\in{\cal N}_s,~t\in{\cal T},\\
            &\hspace{-3mm}D_{nte},\quad\forall n\in{\cal N}\backslash{\cal N}_s,~t\in{\cal T},
        \end{dcases}
    \end{split}\label{TSO_P_balance}\\          
    &(\underline{\gamma}_{ite},\overline{\gamma}_{ite}):\!\quad 0 \le {g}_{ite} \le \overline{g}_{ite},~\forall i \in {\cal{I}}\cup\hat{\cal{I}},~t\in{\cal T}\!,\label{TSO_G_plim}\\
    &(\underline{\delta}_{lte},\overline{\delta}_{lte}):\!\quad -{F}^{\mathrm{max}}_l \le f_{lte} \le {F}^{\mathrm{max}}_l,~\forall {l \in {\cal L}},~t\in{\cal T}\!,\!\!\label{TSO_LineLim}    
\end{align}\label{LLproblem}\end{subequations}
where~$\Xi^{\mathrm{W}}=\{g_{ite}, \theta_{nte}, f_{lte} \ge 0\}$. Objective function \eqref{WM_Obj} maximizes the social welfare under inflexible wholesale demand\footnote{Note that the flexible demand modeled in Eq.~\eqref{eq:flexible_demand} is internalized in the ML problem, i.e. the power utility leverages all flexibility, and the wholesale market thus deals with fixed demand curves, denoted as $d_{nte}$ for nodes in strategically acting state and $D_{nte}$ for others.}, i.e. it minimizes the operating cost. Equation~\eqref{TSO_DSPF} models DC power flows and Eq.~\eqref{TSO_P_balance} enforces the nodal power balance in the transmission system, where $o(l)$ and $r(l)$ denote originating and receiving nodes of line $l$, respectively. The dual variable of Eq.~\eqref{TSO_P_balance} ($\lambda_{nte}$) is then the locational marginal price (LMP). Equations~\eqref{TSO_G_plim}-\eqref{TSO_LineLim} limit generation outputs and transmission line flows.

\section{Solution Method}
To solve the proposed TL model effectively, we convert chance constraints into a deterministic form, derive an equivalent bi-level problem, and develop a C\&CG algorithm variant that deals with bilinear terms across levels.

\vspace{-2mm}
\subsection{Chance constraint reformulation}\label{subsec:CCreform}
As in \cite{kim2020computing}, chance constraints in Eqs.~\eqref{Eq:gen_UB}-\eqref{Eq:gen_LB_new} can be exactly reformulated as the following deterministic constraints:
\begin{subequations}\label{Eq:CC_reform}
\begin{align}    
    \begin{split}
        & \overline{g}_{ite}
        -\alpha_{it} \Big( \hspace{-0.0mm} \sum_{j\in{\cal I}^{\mathrm{R}}_{s}} \hspace{-1.0mm} G^{\mathrm{max}}_j \upsilon_{jte} 
        \!+ \hspace{-1.0mm} \sum_{j\in\hat{\cal I}^{\mathrm{R}}_{s}} \hspace{-1.0mm} g^{\mathrm{max}}_j \upsilon_{jte}\Big)
        \\& \hspace{5.5mm}+ \Phi^{-1}(1\!-\!\eta){\Stdev[\bm{g}_{ite}]} 
        \le {G}_{i}^{\mathrm{max}},~\forall i \in  {\cal I}^{\mathrm{C}}_s, t \in {\cal T},\!\! 
    \end{split}\label{Eq:gen_UB_Conic}\\
    \begin{split}
        & \overline{g}_{ite}
        -\alpha_{it} \Big( \hspace{-0.0mm} \sum_{j\in{\cal I}^{\mathrm{R}}_{s}} \hspace{-1.0mm} G^{\mathrm{max}}_j \upsilon_{jte} 
        \!+ \hspace{-1.0mm} \sum_{j\in\hat{\cal I}^{\mathrm{R}}_{s}} \hspace{-1.0mm} g^{\mathrm{max}}_j \upsilon_{jte}\Big)
        \\& \hspace{5.5mm}+ \Phi^{-1}(1\!-\!\eta){\Stdev[\bm{g}_{ite}]} 
        \le {g}_{i}^{\mathrm{max}},~\forall i \in  \hat{\cal I}^{\mathrm{C}}_s, t \in {\cal T}, 
    \end{split}\label{Eq:gen_UB_New_Conic}\\
% \end{align}
% \begin{align}
    \begin{split}
        & \overline{g}_{ite}
        -\alpha_{it} \Big( \hspace{-0.0mm} \sum_{j\in{\cal I}^{\mathrm{R}}_{s}} \hspace{-1.0mm} G^{\mathrm{max}}_j \upsilon_{jte} 
        \!+ \hspace{-1.0mm} \sum_{j\in\hat{\cal I}^{\mathrm{R}}_{s}} \hspace{-1.0mm} g^{\mathrm{max}}_j \upsilon_{jte}\Big)
        \\& \hspace{5.5mm}- \Phi^{-1}(1\!-\!\eta){\Stdev[\bm{g}_{ite}]} 
        \ge {G}_{i}^{\mathrm{min}},~\forall i \in  {\cal I}^{\mathrm{C}}_s, t \in {\cal T}, \!
    \end{split}\label{Eq:gen_LB_Conic}\\
    \begin{split}
        & \overline{g}_{ite}
        -\alpha_{it} \Big( \hspace{-0.0mm} \sum_{j\in{\cal I}^{\mathrm{R}}_{s}} \hspace{-1.0mm} G^{\mathrm{max}}_j \upsilon_{jte} 
        \!+ \hspace{-1.0mm} \sum_{j\in\hat{\cal I}^{\mathrm{R}}_{s}} \hspace{-1.0mm} g^{\mathrm{max}}_j \upsilon_{jte}\Big)
        \\& \hspace{5.5mm}- \Phi^{-1}(1\!-\!\eta){\Stdev[\bm{g}_{ite}]} 
        \ge \Gamma_i {g}_{i}^{\mathrm{max}},~\forall i \in  \hat{\cal I}^{\mathrm{C}}_s, t \in {\cal T}, \hspace{-10mm}
    \end{split}\label{Eq:gen_LB_New_Conic}\\
    & \Stdev[\bm{g}_{ite}] \!=\! \alpha_{it}\!\sqrt{\sum_{j\in{\cal{I}}_{s(i)}^{\mathrm{R}}}({{G}^{\mathrm{max}}_{j}} \sigma_{jte})^2 \!+\!\! \sum_{j\in\hat{\cal{I}}_{s(i)}^{\mathrm{R}}}({{g}^{\mathrm{max}}_{j}} \sigma_{jte})^2 }, \label{Eq:stdev}
\end{align}\end{subequations}
where $\Phi^{-1}(\cdot)$ is the inverse cumulative distribution function of the Gaussian distribution and $\Stdev[\cdot]$ is a standard deviation operator. Eq.~\eqref{Eq:stdev} assumes independence of random variables $\{\bm{g}_{ite} ~| ~\forall i\!\in \!{\cal I}_s^{\mathrm{R}}\!\cup\!\hat{\cal I}_s^{\mathrm{R}}\}$ across time $t$ and representative days $e$.

\vspace{-2mm}
\subsection{Equivalent bi-level problem}\label{subsec:bilevelreform}
\vspace{-1mm}
Since the LL problem in Eq.~\eqref{LLproblem} is a linear problem, it is guaranteed to have a strict complementary solution by the \textit{Goldman-Tucker theorem} \cite{goldman1956theory}. Hence, we can apply the strong duality theorem to the LL problem and  convert it into a set of constraints, resulting in an equivalent ML and LL problem. This equivalent problem makes it possible to convert the proposed TL problem into an equivalent bi-level problem. 

First, we formulate the dual problem of Eq.~\eqref{LLproblem} for each representative day $e\in{\cal E}$ as:
\begin{subequations}\begin{align}
    \begin{split}
        & \!\underset{\Xi^{\mathrm{DW}}}{\min} O^{\mathrm{DW}}_{e}\coloneqq \sum_{t\in {\cal T}} \Big( \sum_{i \in {\cal I}\cup\hat{\cal I}} \! \overline{g}_{ite} \overline{\gamma}_{ite}
        - \!\!\sum_{n \in {\cal N}} \!D_{nte}^{\mathrm{p}} \lambda_{nte} \!\!\\
        &\hspace{30mm}   + \sum_{l \in {\cal L}} \! \big({F}^{\mathrm{max}}_l \overline{\delta}_{lte}\!+\!{F}^{\mathrm{max}}_l \underline{\delta}_{lte} \big) \Big)
    \end{split}\label{TSO_Obj_dual}\\
    & (g_{ite}^{\mathrm p}):\hspace{2mm}-\underline{\gamma}_{ite}\!+\!\overline{\gamma}_{ite}\!-\!\lambda_{n(i),te} \!=\! - C^{\mathrm g}_i ,~\forall i \in {\cal{I}},t \in {\cal T},\label{DLL_gp}\\
    \begin{split}
        & (f_{lte}^{\mathrm p}):\hspace{2mm}\!-\underline{\delta}_{lte}+\overline{\delta}_{lte}+\xi_{lte} \!-\!\lambda_{r(l),t,e}+\lambda_{o(l),t,e} = 0 ,
        \\&\hspace{55mm}\forall l \in {\cal{L}},t \in {\cal T},
    \end{split}\label{DLL_fl}\hspace{-1mm}\\
    & (\theta_{nte}):\hspace{2mm}-\!\!\!\!\sum_{l \vert o(l) = n}\! \frac{\xi_{lte}}{X_l}\!+\!\!\!\!\!\sum_{l \vert r(l) = n}\! \frac{\xi_{lte}}{X_l} = 0, ~\forall n \in {\cal{N}},t \in {\cal T},\label{DLL_theta}\!
    \end{align}\end{subequations}
where $\Xi^{\mathrm{DW}} = \big\{ \underline{\gamma}_{ite}, \overline{\gamma}_{ite}, \underline{\delta}_{lte}, \overline{\delta}_{lte} \geq 0;  \xi_{lte}, \lambda_{nte} \text{: free} \big\}$.

Then, the proposed TL problem can be equivalently recast as a compact bi-level problem:
\begin{subequations}\label{Eq:bilevel}\begin{align}
    & \text{Eq.}~\eqref{ULproblem}\hspace{31.5mm}:\text{UL problem}
\end{align}
\begin{align}
    \begin{split}
        & \Xi^{\mathrm{U}} \cup \Xi^{\mathrm{W}} \in \arg\big[ 
        \\& \hspace{7mm}\text{Eqs.}~\eqref{Utility_Obj},\eqref{Eq:ramp}\text{\textendash}\eqref{Eq:pdown_ULB}, \eqref{Eq:CC_reform}\hspace{1.5mm} :\text{ML problem}\\    
        & \hspace{7mm}\text{Eqs.}~\eqref{TSO_DSPF}\text{\textendash}\eqref{TSO_LineLim}\hspace{14mm}:\text{LL primal constraints}\\
        & \hspace{7mm}\text{Eqs.}~\eqref{DLL_gp}\text{\textendash}\eqref{DLL_theta}\hspace{14mm}:\text{LL dual constraints}\\
        & \hspace{7mm}O^{\mathrm{W}}_{e}=O^{\mathrm{DW}}_{e},\quad\forall e\in{\cal E}\hspace{1.3mm}:\text{LL strong duality}
        \hspace{3mm}\big].
    \end{split}\label{Eq:combinedMLLL}
\end{align}\end{subequations}
The complete formulation of \eqref{Eq:bilevel} with all constraints detailed is described in Appendix.
It is noteworthy that duality theory is not applicable to the combined ML and LL problem in \eqref{Eq:combinedMLLL} due to the existence of bilinear terms $\tau^{\mathrm{e}} g_{ite},\tau^{\mathrm{c}} g_{i}^{\mathrm{max}}, \lambda_{n(i),t,e} g_{ite}$ in objective functions, and therefore, further conversion of the bi-level problem in \eqref{Eq:bilevel} into an exact single-level form is unattainable. This motivates the use of the C\&CG algorithm for solving \eqref{Eq:bilevel} described in the next subsection. 

\subsection{Column-and-cut generation algorithm}
\begin{figure}[b]
    \centering    
        \includegraphics[width=0.9\columnwidth]{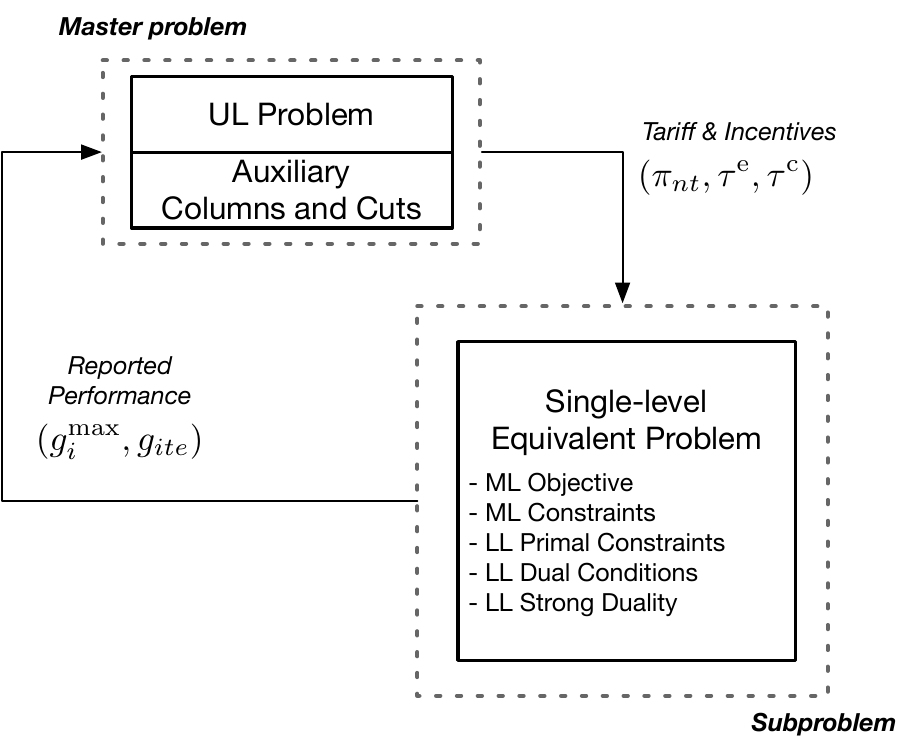}
        \caption{\small Implementation of the  C\&CG algorithm. \vspace{-0mm}}
        \label{fig:CCGstructure}       
\end{figure} 

\begin{figure}[t]
    \centering    
    \includegraphics[width=0.97\columnwidth]{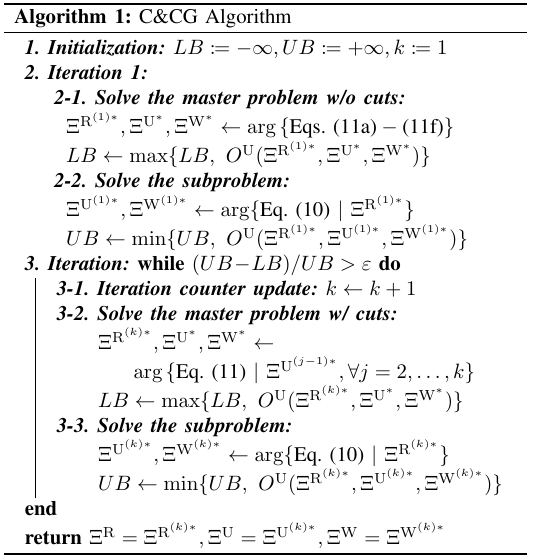}
    \vspace{-0mm}
\end{figure} 

The C\&CG algorithm exploits the primary-secondary (also known as master-slave) structure of Eq.~\eqref{Eq:bilevel} and generates primal cuts and columns over iterations using primal information exchanges between the master- and sub-problems. In our case, as shown in Fig.~\ref{fig:CCGstructure}, the master problem provides policy decisions of the state regulator to the subproblem and receives reported performance (expansion and dispatch decisions) in return. 

The C\&CG algorithm as applied to Eq.~\eqref{Eq:bilevel} is detailed in Algorithm 1. Step 1 initializes the lower and upper bounds of the objective function ($LB,UB$) and the iteration counter ($k$). Step 2-1 obtains initial solutions of the master problem without any cut and update $LB$ and given the UL solution, while Step 2-2 solves a subproblem, saves solutions, and update $UB$. Step 3 first updates the iteration counter (Step 3-1). Then  it obtains the master problem solution  with $k\!-\!1$ sets of cuts generated from the subproblem solutions from all the previous iterations ($1,\dots,k\!-\!1$) and updates $LB$ (Step 3-2). Finally, it solves the subproblem with the solution of the master problem at the current iteration $k$ and updates $UB$ (Step 3-3). This process repeats until the convergence is achieved at a desired tolerance level, $(UB-LB)/UB \le \epsilon$, i.e., the approximation error of the master problem on the subproblem objective $(O^{\mathrm U})$ is within the desired tolerance level.

\subsubsection{Subproblem}
Equation~\eqref{SP} solves the combined ML and LL equivalent problem, while the UL decisions of the current iteration $\Xi^{\mathrm{R}^{(k)*}}$ are given. The subproblem is a bilinear quadratic problem due to $\lambda_{nte} d_{nte}$ and $\lambda_{n(i),t,e} g_{ite}$, and the size of the subproblem does not change over the iterations:
\begin{subequations}\label{SP}\begin{align}          
    \begin{split}             
        &\max_{\Xi^{\mathrm{SP}}} O^{\mathrm{SP}}\!\coloneqq\!\sum_{e\in{\cal E}} \omega_e \Big[
        \hspace{-0.0mm}\sum_{t\in{\cal T}}\sum_{n\in{\cal N}_s}  d_{nte} \pi_{nt}^{(k),*} \hspace{-1mm} 
        + \sum_{t\in{\cal T}}\hspace{-0.5mm}\sum_{i\in\hat{\cal I}^{\mathrm R}\cup{\cal I}^{\mathrm R}}  \hspace{-4mm}\tau^{\mathrm{e}^{(k),*}}\hspace{-1.5mm} g_{ite}   
        \\&\hspace{1.5mm}
        + \hspace{-0.5mm}\sum_{i\in\hat{\cal I}^{\mathrm R}} \tau^{\mathrm{c}^{(k),*}} g^{\mathrm{max}}_i 
        - \hspace{-0.5mm}\sum_{t\in{\cal T}} \Big( \hspace{-0.5mm}\sum_{n\in{\cal N}_s} \hspace{-1.5mm} \lambda_{nte} d_{nte}\hspace{-1mm}
        -  \hspace{-1.5mm}\sum_{i \in {\cal I}\cup\hat{\cal I}} \hspace{-1.5mm}\lambda_{n(i),t,e} {g}_{ite} \Big) 
        \\&\hspace{1.5mm}
        - \sum_{i\in\hat{\cal I}} C^{\mathrm{inv}}_i g^{\mathrm{max}}_{i}
        - \sum_{i\in{\cal I}}\sum_{t\in{\cal T}} C^{\mathrm{g}}_i g_{ite} \Big]
        \label{SP:obj}
    \end{split}
\end{align}
\begin{align}
    & \text{subject to:}\nonumber\\  
    & \text{Eqs.}~\eqref{Eq:ramp}\text{\textendash}\eqref{Eq:pdown_ULB}, \eqref{Eq:CC_reform}:\hspace{9.5mm}\text{ML constraints}\\    
    & \text{Eqs.}~\eqref{TSO_DSPF}\text{\textendash}\eqref{TSO_LineLim}:\hspace{15mm}\text{LL primal constraints}\label{SP:LLconst}\\
    & \text{Eqs.}~\eqref{DLL_gp}\text{\textendash}\eqref{DLL_theta}:\hspace{15mm}\text{LL dual constraints}\label{SP:DLLconst}\\
    & O^{\mathrm{W}}_{e}=O^{\mathrm{DW}}_{e},~\forall e\in{\cal E}:\hspace{4.7mm}\text{LL strong duality}\label{SP:LLduality}
\end{align}\end{subequations}
where $\Xi^{\mathrm{SP}}\coloneqq \Xi^{\mathrm{U}} \cup \Xi^{\mathrm{W}}$.

\subsubsection{Master problem}
Equation~\eqref{MP:obj} minimizes the objective function of state regulator $s$ over variables of all three levels, where Eqs.~\eqref{MP:ULconstraints}\textendash\eqref{MP:LLconst}  define feasible regions of each level. Eqs.~\eqref{MP:LLconst}\textendash\eqref{MP:LLduality} ensure the LL optimum is achieved. In addition, $k-1$ sets of column-and-cut generation constraints are imposed in Eqs.~\eqref{MP:objcut}\textendash\eqref{MP:strongduality}. 
For every C\&CG set, a cut with respect to the ML objective function is generated in Eq.~\eqref{MP:objcut}, with \textit{oracular} knowledge of $\Xi^{\mathrm{U}^{(j)*}}$ given from the subproblem. This allows the master problem to approximate the possible decision-making of the ML and LL problem.
Eqs.~\eqref{MP:DCPF}\textendash\eqref{MP:strongduality} keep auxiliary variables in $\Xi^{\mathrm{Aux}^{(j)}}$ within feasible regions of the LL problem. Note that similarly to the subproblem the master problem is also bilinear quadratic, but the number of auxiliary variables and constraints increase proportionally to the number of iterations:
\begin{subequations}\label{MP}\begin{align}    
	 \begin{split}
        & \min_{\Xi^{\mathrm{MP}}} O^{\mathrm{MP}} \coloneqq O^{\mathrm{R}}_s 
    \end{split}\label{MP:obj}\\
    & \text{subject to:}\nonumber\\
    & \text{Eqs.}~\eqref{Eq:Tariff_budget}\text{\textendash}\eqref{Eq:Radeq}:\hspace{14.8mm}\text{UL constraints}\label{MP:ULconstraints}\\
    & \text{Eqs.}~\eqref{Eq:ramp}\text{\textendash}\eqref{Eq:pdown_ULB}, \eqref{Eq:CC_reform}:\hspace{9.5mm}\text{ML constraints}\label{MP:MLconst}\\
    & \text{Eqs.}~\eqref{TSO_DSPF}\text{\textendash}\eqref{TSO_LineLim}:\hspace{15mm}\text{LL primal constraints}\label{MP:LLconst}\\
    & \text{Eqs.}~\eqref{DLL_gp}\text{\textendash}\eqref{DLL_theta}:\hspace{15mm}\text{LL dual constraints}\label{MP:DLLconst}\\
    & O^{\mathrm{W}}_{e}=O^{\mathrm{DW}}_{e},~\forall e\in{\cal E}:\hspace{4.7mm}\text{LL strong duality}\label{MP:LLduality}\\
    & \text{with the C\&CG constraints updated at every iteration $k$ as:}\hspace{15mm}\nonumber\\
    \begin{split}            
        & \Bigg[\hspace{-0.5mm}
            \sum_{t\in{\cal T}}\sum_{n\in{\cal N}_s}  d_{nte} \pi_{nt}  
            + \sum_{t\in{\cal T}}\hspace{-0.5mm}\sum_{i\in\hat{\cal I}^{\mathrm R}\cup{\cal I}^{\mathrm R}}  \tau^{\mathrm{e}} g_{ite}   
            \\&\hspace{1mm}
            + \hspace{-0.5mm}\sum_{i\in\hat{\cal I}^{\mathrm R}} \tau^{\mathrm{c}} g^{\mathrm{max}}_i  
            - \hspace{-0.5mm}\sum_{t\in{\cal T}} \Big( \hspace{-0.5mm}\sum_{n\in{\cal N}_s} \hspace{-1.5mm} \lambda_{nte} d_{nte}\hspace{-1mm}
            -  \hspace{-1.5mm}\sum_{i \in {\cal I}\cup\hat{\cal I}} \hspace{-1.5mm}\lambda_{n(i),t,e} {g}_{ite} \Big) 
            \\&\hspace{1mm}
            - \sum_{i\in\hat{\cal I}} C^{\mathrm{inv}}_i g^{\mathrm{max}}_{i}
            - \sum_{i\in{\cal I}}\sum_{t\in{\cal T}} C^{\mathrm{g}}_i g_{ite}            
        \hspace{0.5mm}\ge
        \\& \hspace{1.5mm}
            \sum_{t\in{\cal T}}\sum_{n\in{\cal N}_s}  d_{nte} \pi_{nt}  
            + \sum_{t\in{\cal T}}\hspace{-0.5mm}\sum_{i\in\hat{\cal I}^{\mathrm R}\cup{\cal I}^{\mathrm R}}  \tau^{\mathrm{e}} g_{ite}^{(j)}   
            \\&\hspace{1mm}
            + \hspace{-0.5mm}\sum_{i\in\hat{\cal I}^{\mathrm R}} \tau^{\mathrm{c}} g^{\mathrm{max}^{(j-1),*}}_i  \hspace{-2mm}
            - \hspace{-0.5mm}\sum_{t\in{\cal T}} \Big( \hspace{-0.5mm}\sum_{n\in{\cal N}_s} \hspace{-1.5mm} \lambda_{nte}^{(j)} d_{nte}\hspace{-1mm}
            -  \hspace{-1.5mm}\sum_{i \in {\cal I}\cup\hat{\cal I}} \hspace{-1.5mm}\lambda_{n(i),t,e}^{(j)} {g}_{ite}^{(j)} \Big) 
            \\&\hspace{1mm}
            - \sum_{i\in\hat{\cal I}} C^{\mathrm{inv}}_i g^{\mathrm{max}^{(j-1),*}}_{i}
            - \sum_{i\in{\cal I}}\sum_{t\in{\cal T}} C^{\mathrm{g}}_i g_{ite}^{(j)}                
    \end{split}\label{MP:objcut}\\
    & f_{lte}^{(j)} = \frac{1}{X_l} (\theta_{o(l),te}^{(j)}-\theta_{r(l),te}^{(j)}),~\forall {l \in {\cal L}},~t\in{\cal T},\label{MP:DCPF}\\    
    \begin{split}
        &\hspace{-1mm}\sum_{i \in {\cal I}_n\cup\hat{\cal I}_n } \hspace{-1.5mm}{g}_{ite}^{(j)}  + \hspace{-1mm}\sum_{l \vert r(l) = n } \hspace{-1.5mm} f_{lte}^{(j)} \!-\hspace{-1mm}\sum_{l \vert o(l) = n } \hspace{-1.5mm} f_{lte}^{(j)} 
        \\&\hspace{20mm}=\begin{dcases}
            &{\hspace{-3mm}d_{nte},\quad\forall n\in{\cal N}_s,~t\in{\cal T},}\\
            &{\hspace{-3mm}D_{nte},\quad\forall n\in{\cal N}\backslash{\cal N}_s,~t\in{\cal T},}
        \end{dcases}
    \end{split}\label{MP:Pbalance}\\
    & 0 \le {g}_{ite}^{(j)} \le \overline{g}^{(j-1),*}_{ite},~\forall i \in {\cal{I}}\cup\hat{\cal{I}},~t\in{\cal T}\!,\label{MP:Tglim}\\    
    &-{F}^{\mathrm{max}}_l \le f_{lte}^{(j)} \le {F}^{\mathrm{max}}_l,~\forall {l \in {\cal L}},~t\in{\cal T}\!,\!\!\label{MP:Tllim}\\
    & -\underline{\gamma}_{ite}^{(j)}+\overline{\gamma}_{ite}^{(j)}-\lambda_{n(i),t,e}^{(j)} = - C^{\mathrm g}_i ,\quad\forall i \in {\cal{I}},~t\in{\cal T},\label{MP:DLL_gp}\\
    & \!-\!\underline{\delta}_{lte}^{(j)}\!+\!\overline{\delta}_{lte}^{(j)}\!+\!\xi_{lte}^{(j)} \!-\!\lambda_{r(l),t,e}^{(j)}\!+\!\lambda_{o(l),t,e}^{(j)} \!=\! 0 ,~\forall l \in {\cal{L}},~t\in{\cal T},\label{MP:DLL_fl}\hspace{-1mm}\\
    & -\!\!\!\!\sum_{l \vert o(l) = n}\! \frac{\xi_{lte}^{(j)}}{X_l}~+\!\!\!\!\sum_{l \vert r(l) = n}\! \frac{\xi_{lte}^{(j)}}{X_l} = 0, \quad\forall n \in {\cal{N}},~t\in{\cal T},\label{MP:DLL_theta}\\
    \begin{split}    
        & \sum_{t\in{\cal T}} \sum_{i \in {\cal I}\cup\hat{\cal I}} -C_i^{\mathrm{g}} {g}_{ite}^{(j)} 
        =  \sum_{t\in {\cal T}} \Big( \sum_{i \in {\cal I}\cup\hat{\cal I}} \! \overline{g}^{(j-1),*}_{ite} \overline{\gamma}_{ite}^{(j)}        
        \\&\hspace{10mm}   
        - \!\!\sum_{n \in {\cal N}} \!D_{nte} \lambda_{nte}^{(j)} 
        + \sum_{l \in {\cal L}} \! \big({F}^{\mathrm{max}}_l \overline{\delta}_{lte}^{(j)}\!+\!{F}^{\mathrm{max}}_l \underline{\delta}_{lte}^{(j)} \big) \Big) \!\!
    \end{split}\label{MP:strongduality}\\
    & \Bigg],\quad\forall e\in{\cal E}, ~j = 2,\dots, k, \nonumber
\end{align}\end{subequations}
where 
{ MP decision variable set $\Xi^{\mathrm{MP}}$ is defined as the union of original variable set $\Xi^{\mathrm{R}}\cup \Xi^{\mathrm{U}} \cup \Xi^{\mathrm{W}}$ and auxiliary variable set $\Xi^{\mathrm{Aux}^{(j)}}$ for each C\&CG iteration $j$, i.e. 
}
$\Xi^{\mathrm{MP}}\coloneqq (\Xi^{\mathrm{R}}\cup \Xi^{\mathrm{U}} \cup \Xi^{\mathrm{W}})\cup \bigcup_{j=2}^{k}\Xi^{\mathrm{Aux}^{(j)}}$ and $\Xi^{\mathrm{Aux}^{(j)}}\coloneqq \{g_{ite}^{(j)}, \theta_{nte}^{(j)}, f_{lte}^{(j)}, \xi_{lte}^{(j)}, \lambda_{nte}^{(j)}:\text{free};~ \underline{\gamma}_{ite}^{(j)}, \overline{\gamma}_{ite}^{(j)}, \underline{\delta}_{lte}^{(j)}, \overline{\delta}_{lte}^{(j)} \ge 0 \}$.

\subsubsection{Optimality and convergence}
{
Note that since the master- and sub-problems in \eqref{SP}\textendash\eqref{MP} are bilinear quadratic programming problems, global optimality cannot be guaranteed by the off-the-shelf NLP solvers. The convergence of the proposed C\&CG algorithm largely depends on the solution quality of the master-/sub-problems (that is, how close the obtained optimum is from the global optimum) and an initialization point.
To overcome this issue, in literature, multi-start algorithm \cite{KNITROmulti}, convex envelops \cite{gurobi2020manual}, and binary expansion \cite{pereira2005strategic} are commonly used to achieve quasi-global optimum and improve the solution quality.  As explained in Section~\ref{Sec:casestudy},   we use a solver with a multi-start algorithm to improve convergence and solution quality. 
}

\section{Case Study}\label{Sec:casestudy}
\begin{figure}[t]
    \centering    
    \vspace{-1mm}
        \includegraphics[width=0.82\columnwidth]{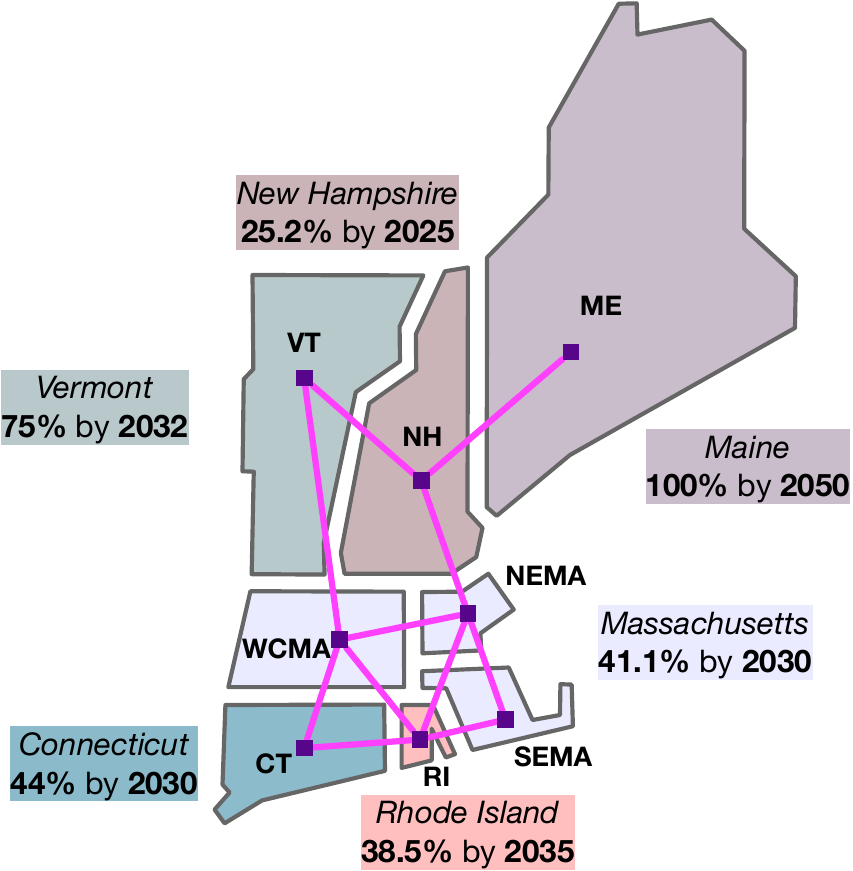}
        \caption{\small An 8-zone ISO-NE  system with RPS goals as of 01/01/2020. \vspace{-3mm}}
        \label{fig:ISONE_8zone}       
\end{figure} 

We use the 8-zone ISO New England (ISO-NE) test system \cite{krishnamurthy20168}, covering six states with the RPS goals as in Fig.~\ref{fig:ISONE_8zone}. Table~\ref{table:gencap} and  Table~\ref{table:genspec} report the current generation mix   and cost characteristics of candidate generators. Expansion decisions of non-strategic states are computed based on the model in \cite{munoz2014engineering}. Similarly to \cite{7464891}, we use a hierarchical clustering algorithm \cite{pitt2000applications} to identify 5 representative days from historical hourly demand and renewable generation forecast data collected from the ISO-NE data library \cite{isone2019final,van2018state}. All fossil-fueled generators are assumed to be committed and the renewable forecast error is zero-mean and has $\sigma_{ite}\!=\!0.1$ for wind and $\sigma_{ite}\!=\!0.2$ for solar, while $\eta\!=\!0.03$. Generation resources in non-strategic  states are assumed to offer their full capacity to the market at their marginal cost. Using the investment recovery period of  10 years and a discount rate of 5\%,  $C^{\mathrm{inv}}_i$ is prorated on a daily basis using a net present value approach. Affine control parameters are set as predefined values $\alpha_{it}=1/\mathrm{card}({\cal I}_s^{\mathrm C}\cup\hat{\cal I}_s^{\mathrm C})$ \cite{jaleeli1992understanding,makarov2009operational}. Demand-side parameters in Eq.~\eqref{eq:flexible_demand} are $N=0.25$ and $M_{nte}=0.25 D_{nte}+20$ \cite{samadi2010optimal}, while peak TOU hours are set as 13:00-21:00, \cite{Eversource}. All models are implemented using Julia v1.3/JuMP v0.21 and solved by the KNITRO solver v12.1 on an Intel Xeon 2.6 GHz CPU with 124GB of memory. The solution procedure is initialized with 300 different starting points, which improves computational performance and overcomes  potential issues arising from inborn local optimality of the bilinear quadratic problems. The average computation time of the  C\&CG algorithm was 97.4 hours with at most 5 iterations, which is acceptable for long-term planning problems when there is no pressure to instantly enforce the optimal decisions. The input data and code are available in \cite{julia_code_gist}.

\begin{table}[t]
        \centering
        \captionsetup{justification=centering, labelsep=period, font=footnotesize, textfont=sc}
        \caption{Generation capacity by state and type [MW]\vspace{-0mm}}        
        \begin{tabular}{L{14mm}  R{7.5mm} R{7.5mm} R{7.5mm} R{7.5mm} R{7.5mm} R{7.5mm}}    
            \toprule
            &\multicolumn{1}{c}{ME}&\multicolumn{1}{c}{NH}&\multicolumn{1}{c}{VT}&\multicolumn{1}{c}{MA}&\multicolumn{1}{c}{CT}&\multicolumn{1}{c}{RI}\\
            \midrule
            Wind&221.2&140.5&39.0&681.7&132.5&85.0\\
            Solar&41.4&83.8&306.3&1871.3&464.3&116.7\\
            Nuclear&\textendash&1244.0&620.2&684.7&2116.0&\textendash\\
            Coal&311.8&95.4&\textendash&144.4&744.4&1099.5\\
            Oil&1146.9&400.2&\textendash&1111.7&2212.8&435.0\\
            Natural Gas&3862.7&508.0&\textendash&2249.6&621.4&3491.6\\
            \bottomrule
        \end{tabular}
        \label{table:gencap}              
    \vspace{-0mm}
    \vspace{3mm}
        \centering
        \captionsetup{justification=centering, labelsep=period, font=footnotesize, textfont=sc}
        \caption{Investment and operation costs of candidate generators \cite{EIA_cost}\vspace{-0mm}}   
        \begin{tabular}{L{29mm}  P{14mm} P{14mm} P{14mm} }    
            \toprule
            \multirow{2}{*}{}&\multicolumn{1}{c}{Wind}&\multicolumn{1}{c}{Solar}&\multicolumn{1}{c}{Natural Gas}\\
            \midrule
            Investment Cost, \$/kW&1630&2434&895\\
            Operating Cost, \$/MWh&1.1&0.4&20.0\\
            \bottomrule
        \end{tabular}
        \label{table:genspec}              
    \vspace{-0mm}
\end{table}

\subsection{NH Case: investment, policy, cost decisions}\label{sec:case_nh}

\begin{figure}[t!]
    \centering    
    \vspace{-1mm}
        \includegraphics[width=1\columnwidth]{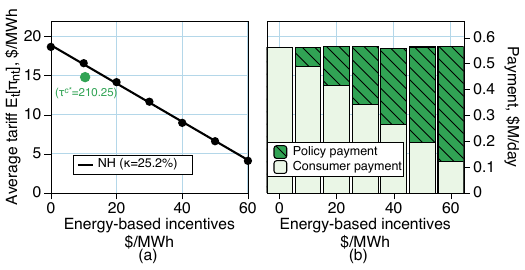}
        \vspace{-8mm}
        \caption{\small  (a) Relationship between the optimal incentives and electricity tariff  for (i) the co-optimization of energy- and capacity-based incentives and tariff (green circle) and (ii) for the co-optimization of the energy-based incentive and tariff, and (b) comparison between the consumer payment and the cost of energy-based incentives. \vspace{0mm}}
        \label{fig:NH_EPayment}
    \centering    
        \includegraphics[width=1\columnwidth]{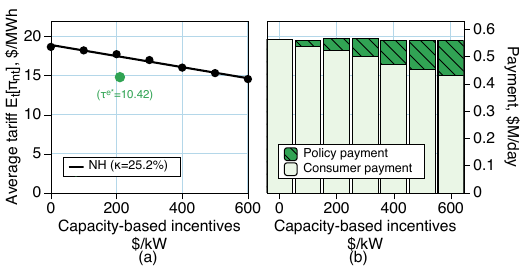}
        \vspace{-8mm}
        \caption{\small  (a) Relationship between the optimal incentives and electricity tariff for (i) the co-optimization of energy- and capacity-based incentives and tariff (green circle) and (ii) for the co-optimization of the capacity-based incentive and tariff, and (b) comparison between the consumer payment and the cost of capacity-based incentives. \vspace{-3mm}}
        \label{fig:NH_CPayment}
\end{figure}

\begin{figure}[p]
    \centering    
    \vspace{-0mm}
        \includegraphics[width=1.0\columnwidth]{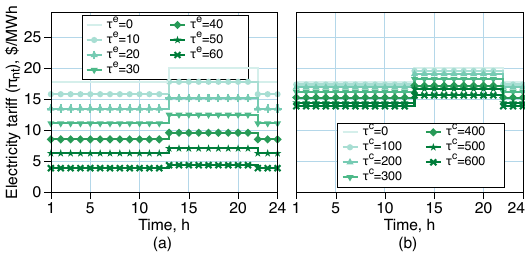}
        \vspace{-7mm}
        \caption{\small  Optimal time-of-use (TOU) electricity tariffs for  (a) energy-based and (b) capacity-based incentives. The darker the color, the higher the value of incentives.  \vspace{1mm}}
        \label{fig:TOUs}           
    \centering    
    \vspace{-0mm}
        \includegraphics[width=\columnwidth]{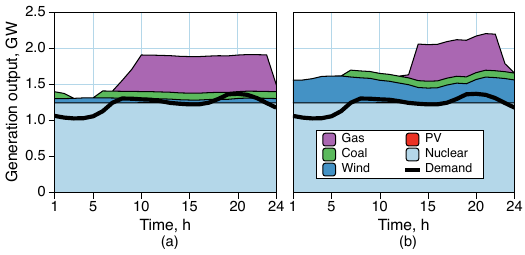} 
        \vspace{-7mm}
        \caption{\small Dispatch of generators and demand in NH: (a) before and (b) after implementing RPS goals. \vspace{0mm}}
        \label{fig:NH_dispatch_joint} 
    \centering    
    \vspace{-1mm}
        \includegraphics[width=\columnwidth]{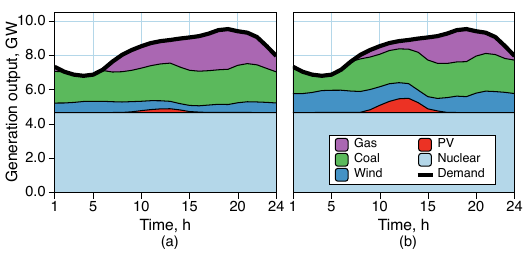}
        \vspace{-7mm}
        \caption{\small Dispatch of generators and demand in the entire ISO-NE system: (a) before and (b) after implementing RPS goals. \vspace{0mm}}
        \label{fig:ISONE_dispatch_joint}       
    \centering    
    \vspace{-0mm}
        \includegraphics[width=0.98\columnwidth]{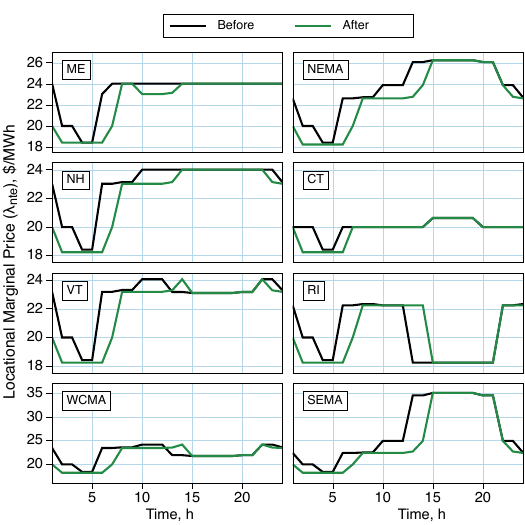}
        \vspace{-3mm}
        \caption{\small Locational marginal prices (LMPs) in the ISO-NE test system before and after implementing RPS goals. \vspace{-0mm}}
        \label{fig:LMP}   
\end{figure}

This section focuses on the NH case because it has the earliest RPS goal among other states. The RPS goal is achieved by installing 602 MW of wind generation resources and requires no enhancement to the fossil-fueled mix (due to a high operating cost). 
{ 
Fig.~\ref{fig:NH_EPayment}(a) and \ref{fig:NH_CPayment}(a) present the optimal combinations of incentives and average tariff ($\mathbb{E}_t [\pi_{nt}]$), which is computed for each case based on the hourly TOU values in  Fig.~\ref{fig:TOUs}, to support the RPS goal in NH. To obtain these results, we carry out two types of simulations: (i) the simultaneous implementation of energy- and capacity-based incentives, i.e. co-optimization of three regulatory decision variables ($\pi_{nt}, \tau^{\mathrm{e}}, \tau^{\mathrm{c}}$), and (ii) the case when policymakers wish to implement either one of the incentives, i.e. co-optimization of one of the incentives ($\tau^{\mathrm{e}}$ or $ \tau^{\mathrm{c}}$) and the tariff ($\pi_{nt}$). In the first case, a unique combination of three regulatory decision variables ($\pi_{nt}, \tau^{\mathrm{e}}, \tau^{\mathrm{c}}$) is determined by the proposed TL model as shown by green dots ($\mathbb{E}_t [\pi_{nt}]\!=\!\$14.8/\mathrm{MWh}, \tau^{\mathrm{e}}\!=\!\$10.42\mathrm{MWh}, \tau^{\mathrm{c}}\!=\!\$201.25\mathrm{kW}$) in Fig.~\ref{fig:NH_EPayment}(a) and \ref{fig:NH_CPayment}(a). In the second case, both the energy- and capacity-based incentives linearly depend on the average tariff.} Indeed, as the average tariff reduces, which in turn leads to a lower consumer payment, greater energy and capacity-based incentives are required to achieve the same RPS goal. Additionally,  Fig.~\ref{fig:NH_EPayment}(b) and \ref{fig:NH_CPayment}(b) summarize the utility revenue from both the incentives and tariffs.  Notably, as Fig.~\ref{fig:NH_EPayment} and \ref{fig:NH_CPayment} show,  the energy-based incentive of  \$20/MWh and the capacity-based incentive of \$600/{kW} yield the same value of the average tariff to achieve the NH RPS goal. 


Fig.~\ref{fig:NH_dispatch_joint}\textendash\ref{fig:LMP} compare the average dispatch and LMPs across the entire ISO NE test system before and after achieving RPS goals. Due to the increased wind capacity, NH produces more wind power and  reduces the usage of fossil-fueled resources (e.g. natural gas). Despite this reduction, the amount of power exported from NH to other states increases as shown in Fig.~\ref{fig:NH_dispatch_joint}(b). As a result of these dispatch changes,  LMPs in the system   decrease by $\approx$2.5\% on average as shown in Fig.~\ref{fig:LMP}.

Next, to analyze the sensitivity of the optimal policy decisions in Fig.~\ref{fig:NH_EPayment} and \ref{fig:NH_CPayment}, the current NH RPS goal is compared to the lower and higher targets, which are set to 85\% and 115\% of the original RPS goal. As Fig.~\ref{fig:pi_tau_joint} shows, the linear relationship between the average tariff and the energy- and capacity-based incentives perseveres. Furthermore, the slope in Fig.~\ref{fig:pi_tau_joint} becomes steeper for the high RPS scenario, thus indicating that a higher RPS goal requires either a greater average tariff for consumers or greater incentives.

\subsection{Comparison with the MA and ME cases}\label{Subsec:RPS_Sens}

\begin{figure}[t!]
    \centering
    \includegraphics[scale=1]{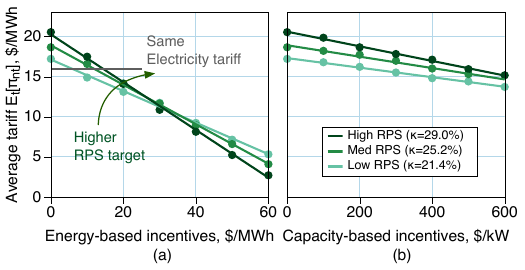}
        \vspace{-7mm}
        \caption{\small Sensitivity of the relationship between the optimal incentives and electricity tariff to an RPS goal. \vspace{2mm}}
        \label{fig:pi_tau_joint}     
    \centering
    \includegraphics[scale=1]{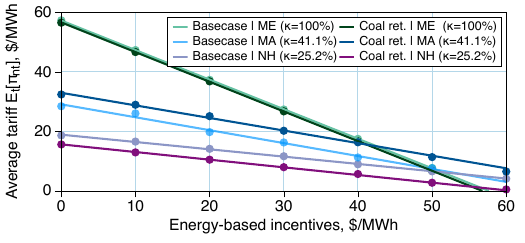}
        \vspace{-7mm} 
        \caption{\small Sensitivity of the relationship between the optimal incentives and electricity tariff in ME, MA and NH to coal retirements. \vspace{-3mm}}
        \label{fig:Ret_multi}  
\end{figure} 

The results attained in Section~\ref{sec:case_nh} can be extended to the MA and ME cases, which have 41.1 and 100\% RPS goals as in Fig.~\ref{fig:ISONE_8zone}. In the following, we also compared the case with no retirement (base case) and the case with the full retirement of coal-fired resources. As  Fig.~\ref{fig:Ret_multi} shows, the linear relationship between the average tariff and the energy- and capacity-based incentives persevere for all scenarios. However, a more aggressive RPS goal leads to a steeper slope. 

The difference between the base and retirement cases manifests itself differently for each state. There is almost no difference in the most aggressive RPS case of ME, where 311.8MW of coal-fired resources can be retired with no implications on the optimal policies. This is due to the fact that the state of ME has a $100\%$ RPS goal, and thus, coal-fired resources would not be dispatched due to a relatively high operating cost once the renewable generation expansion is made. In the case of MA, the retirement will increase the policy incentives needed to  support the RPS goal for the same average tariff.  On the other hand, in the case of NH, coal retirements reduce both the incentives and average tariffs needed to support their modest RPS goal. This is due to the replacement of coal-fired resources in the optimal dispatch with relatively cheap wind and gas producer.

\section{Conclusion}\label{Sec:conclusion}

This paper proposed a TL framework to support a renewable policy decision-making process of a state regulator that internalizes conflicting interests of different stakeholders. The resulting problem is a computationally complex TL bilinear optimization problem and, thus, the C\&CG algorithm is implemented to solve it efficiently. The case study compared the optimal  policies supporting the implementation of RPS goals. The numerical results obtained from the the 8-zone ISO NE test system demonstrated that a more ambitious RPS target requires either an increased electricity tariff (consumer payment) or higher RES incentives (policy support) under the same retirement scenario. Furthermore, our numerical results demonstrate that there is a linear relationship between the RES incentives and  the average  electricity tariff to implement given RPS goals. However, the retirement policy of fossil-fueled generators such as coal-fired power plants can increase or decrease the total cost (incentives and tariffs hikes) of  achieving the same given RPS goals, depending on the current generation mix of the jurisdiction.

\bibliographystyle{IEEEtran}
\bibliography{Trilevel.bib}

\begin{thebibliography}{10}
\providecommand{\url}[1]{#1}
\csname url@samestyle\endcsname
\providecommand{\newblock}{\relax}
\providecommand{\bibinfo}[2]{#2}
\providecommand{\BIBentrySTDinterwordspacing}{\spaceskip=0pt\relax}
\providecommand{\BIBentryALTinterwordstretchfactor}{4}
\providecommand{\BIBentryALTinterwordspacing}{\spaceskip=\fontdimen2\font plus
\BIBentryALTinterwordstretchfactor\fontdimen3\font minus
  \fontdimen4\font\relax}
\providecommand{\BIBforeignlanguage}[2]{{%
\expandafter\ifx\csname l@#1\endcsname\relax
\typeout{** WARNING: IEEEtran.bst: No hyphenation pattern has been}%
\typeout{** loaded for the language `#1'. Using the pattern for}%
\typeout{** the default language instead.}%
\else
\language=\csname l@#1\endcsname
\fi
#2}}
\providecommand{\BIBdecl}{\relax}
\BIBdecl

\bibitem{EIA_CO2}
\BIBentryALTinterwordspacing
{U.S. Energy Information Administration}, ``{How much of U.S. carbon dioxide
  emissions are associated with electricity generation?}'' 2019. [Online].
  Available: \url{https://www.eia.gov/tools/faqs/faq.php?id=77&t=11}
\BIBentrySTDinterwordspacing

\bibitem{hogan2015cleaner}
W.~W. Hogan, ``A cleaner energy system: Renewable energy and electricity market
  design [in my view],'' \emph{IEEE Power and Energy Magazine}, vol.~13, no.~4,
  pp. 112--109, 2015.

\bibitem{aldy2010designing}
J.~E. Aldy, A.~J. Krupnick, R.~G. Newell, I.~W. Parry, and W.~A. Pizer,
  ``Designing climate mitigation policy,'' \emph{Journal of Economic
  Literature}, vol.~48, no.~4, pp. 903--34, 2010.

\bibitem{fischer2008environmental}
C.~Fischer and R.~G. Newell, ``Environmental and technology policies for
  climate mitigation,'' \emph{Journal of environmental economics and
  management}, vol.~55, no.~2, pp. 142--162, 2008.

\bibitem{palmer2011federal}
K.~Palmer, A.~Paul, M.~Woerman, and D.~C. Steinberg, ``Federal policies for
  renewable electricity: Impacts and interactions,'' \emph{Energy Policy},
  vol.~39, no.~7, pp. 3975--3991, 2011.

\bibitem{Soren2019}
S.~T. Anderson, I.~Marinescu, and B.~Shor, ``{Can Pigou at the Polls Stop Us
  Melting the Poles?}'' National Bureau of Economic Research, Working Paper
  26146, August 2019.

\bibitem{Dsire}
\BIBentryALTinterwordspacing
{DsireUSA}, ``{Database of State Incentives for Renewables \& Efficiency},''
  2020. [Online]. Available:
  \url{https://programs.dsireusa.org/system/program?fromSir=0&state=MA}
\BIBentrySTDinterwordspacing

\bibitem{murdock2019renewables}
H.~E. Murdock \emph{et~al.}, ``Renewables 2019 global status report,'' 2020.

\bibitem{ajadi2019global}
T.~Ajadi \emph{et~al.}, ``Global trends in renewable energy investment 2019,''
  2019.

\bibitem{huntington2017revisiting}
S.~C. Huntington, P.~Rodilla, I.~Herrero, and C.~Batlle, ``Revisiting support
  policies for res-e adulthood: Towards market compatible schemes,''
  \emph{Energy Policy}, vol. 104, pp. 474--483, 2017.

\bibitem{newbery2018market}
D.~Newbery, M.~G. Pollitt, R.~A. Ritz, and W.~Strielkowski, ``Market design for
  a high-renewables european electricity system,'' \emph{Renewable and
  Sustainable Energy Reviews}, vol.~91, pp. 695--707, 2018.

\bibitem{ozdemir2019capacity}
{\"O}.~{\"O}zdemir, B.~F. Hobbs, M.~van Hout, and P.~R. Koutstaal, ``Capacity
  vs energy subsidies for promoting renewable investment: Benefits and costs
  for the eu power market,'' \emph{Energy Policy}, p. 111166, 2019.

\bibitem{nicolini2017renewable}
M.~Nicolini and M.~Tavoni, ``Are renewable energy subsidies effective? evidence
  from europe,'' \emph{Renewable and Sustainable Energy Reviews}, vol.~74, pp.
  412--423, 2017.

\bibitem{bienstock2014chance}
D.~Bienstock, M.~Chertkov, and S.~Harnett, ``Chance-constrained optimal power
  flow: Risk-aware network control under uncertainty,'' \emph{Siam Review},
  vol.~56, no.~3, pp. 461--495, 2014.

\bibitem{ruiz2015robust}
C.~Ruiz and A.~J. Conejo, ``Robust transmission expansion planning,''
  \emph{European Journal of Operational Research}, vol. 242, no.~2, pp.
  390--401, 2015.

\bibitem{duarte2020multi}
J.~L.~R. Duarte, N.~Fan, and T.~Jin, ``Multi-process production scheduling with
  variable renewable integration and demand response,'' \emph{European Journal
  of Operational Research}, vol. 281, no.~1, pp. 186--200, 2020.

\bibitem{dvorkin2017co}
Y.~Dvorkin, R.~Fernandez-Blanco, Y.~Wang, B.~Xu, D.~S. Kirschen,
  H.~Pand{\v{z}}i{\'c}, J.-P. Watson, and C.~A. Silva-Monroy, ``Co-planning of
  investments in transmission and merchant energy storage,'' \emph{IEEE
  Transactions on Power Systems}, vol.~33, no.~1, pp. 245--256, 2017.

\bibitem{moreira2014adjustable}
A.~Moreira, A.~Street, and J.~M. Arroyo, ``An adjustable robust optimization
  approach for contingency-constrained transmission expansion planning,''
  \emph{IEEE Transactions on Power Systems}, vol.~30, no.~4, pp. 2013--2022,
  2014.

\bibitem{Jacobs2018}
B.~Jacobs, ``The marginal cost of public funds is one at the optimal tax
  system,'' \emph{Int Tax Public Finance}, vol.~25, pp. 883–--912, 2018.

\bibitem{BENTO2018}
A.~M. Bento, M.~R. Jacobsen, and A.~A. Liu, ``Environmental policy in the
  presence of an informal sector,'' \emph{Journal of Environmental Economics
  and Management}, vol.~90, pp. 61--77, 2018.

\bibitem{Barrios2013}
S.~Barrios, J.~Pycroft, and B.~Saveyn, ``The marginal cost of public funds in
  the eu: the case of labour versus green taxes,'' \emph{European Commission
  Working Paper Series}, vol. N.35 - 2013, 2013.

\bibitem{samadi2010optimal}
P.~Samadi, A.-H. Mohsenian-Rad, R.~Schober, V.~W. Wong, and J.~Jatskevich,
  ``Optimal real-time pricing algorithm based on utility maximization for smart
  grid,'' in \emph{2010 First IEEE International Conference on Smart Grid
  Communications}.\hskip 1em plus 0.5em minus 0.4em\relax IEEE, 2010, pp.
  415--420.

\bibitem{jaleeli1992understanding}
N.~Jaleeli, L.~S. VanSlyck, D.~N. Ewart, L.~H. Fink, and A.~G. Hoffmann,
  ``Understanding automatic generation control,'' \emph{IEEE transactions on
  power systems}, vol.~7, no.~3, pp. 1106--1122, 1992.

\bibitem{6936940}
{H. Pandzic et al.}, ``Near-optimal method for siting and sizing of distributed
  storage,'' \emph{IEEE Tran. Pwr. Syst.}, vol.~30, no.~5, Sep. 2015.

\bibitem{kim2020computing}
J.~Kim, R.~Mieth, and Y.~Dvorkin, ``Computing a strategic decarbonization
  pathway: A chance-constrained equilibrium problem,'' \emph{arXiv preprint
  arXiv:2002.08298}, 2020.

\bibitem{goldman1956theory}
A.~J. Goldman and A.~W. Tucker, ``Theory of linear programming,'' \emph{Linear
  inequalities and related systems}, vol.~38, pp. 53--97, 1956.

\bibitem{KNITROmulti}
\BIBentryALTinterwordspacing
{Artelys Knitro}, ``{User's Manual - Multistart Algorithm},'' 2020. [Online].
  Available:
  \url{https://www.artelys.com/docs/knitro/2_userGuide/multistart.html}
\BIBentrySTDinterwordspacing

\bibitem{gurobi2020manual}
\BIBentryALTinterwordspacing
Gurobi, ``Gurobi optimizer reference manual,'' 2020. [Online]. Available:
  \url{https://www.gurobi.com/wp-content/plugins/hd_documentations/documentation/9.1/refman.pdf}
\BIBentrySTDinterwordspacing

\bibitem{pereira2005strategic}
M.~V. Pereira, S.~Granville, M.~H. Fampa, R.~Dix, and L.~A. Barroso,
  ``Strategic bidding under uncertainty: a binary expansion approach,''
  \emph{IEEE Transactions on Power Systems}, vol.~20, no.~1, pp. 180--188,
  2005.

\bibitem{krishnamurthy20168}
D.~Krishnamurthy, W.~Li, and L.~Tesfatsion, ``An 8-zone test system based on
  iso new england data: Development and application,'' \emph{IEEE Trans. Pwr.
  Syst.}, vol.~31, no.~1, pp. 234--246, 2016.

\bibitem{munoz2014engineering}
F.~D. Munoz, B.~F. Hobbs, J.~L. Ho, and S.~Kasina, ``An engineering-economic
  approach to transmission planning under market and regulatory uncertainties:
  Wecc case study,'' \emph{IEEE Transactions on Power Systems}, vol.~29, no.~1,
  pp. 307--317, 2014.

\bibitem{7464891}
Y.~{Dvorkin}, R.~{Fernández-Blanco}, D.~S. {Kirschen}, H.~{Pandžić},
  J.~{Watson}, and C.~A. {Silva-Monroy}, ``Ensuring profitability of energy
  storage,'' \emph{IEEE Transactions on Power Systems}, vol.~32, no.~1, pp.
  611--623, 2017.

\bibitem{pitt2000applications}
B.~Pitt, ``Applications of data mining techniques to electric load profiling,''
  Ph.D. dissertation, University of Manchester, 2000.

\bibitem{isone2019final}
\BIBentryALTinterwordspacing
``{ISONE 2019 PV Forecast},'' 2019. [Online]. Available:
  \url{https://www.iso-ne.com/static-assets/documents/2019/04/final-2019-pv-forecast.pdf}
\BIBentrySTDinterwordspacing

\bibitem{van2018state}
\BIBentryALTinterwordspacing
G.~van Welie, ``Iso new england - state of the grid: 2018,'' 2018. [Online].
  Available:
  \url{https://www.iso-ne.com/static-assets/documents/2018/02/02272018_pr_presentation_state-of-the-grid_2018.pdf}
\BIBentrySTDinterwordspacing

\bibitem{makarov2009operational}
Y.~V. Makarov, C.~Loutan, J.~Ma, and P.~De~Mello, ``Operational impacts of wind
  generation on california power systems,'' \emph{IEEE transactions on power
  systems}, vol.~24, no.~2, pp. 1039--1050, 2009.

\bibitem{Eversource}
\BIBentryALTinterwordspacing
{Eversource}, ``{Electric Tariffs and Rules},'' 2019. [Online]. Available:
  \url{https://www.eversource.com/content/ct-c/residential/my-account/billing-payments/about-your-bill/rates-tariffs/}
\BIBentrySTDinterwordspacing

\bibitem{julia_code_gist}
\BIBentryALTinterwordspacing
``{Code Supplement for Strategic Policy-making for Renewable Energy Support: A
  Tri-level Optimization Approach}.'' [Online]. Available:
  \url{https://github.com/jipkim/Trilevel}
\BIBentrySTDinterwordspacing

\bibitem{EIA_cost}
\BIBentryALTinterwordspacing
{US Energy Information Administration}, ``{Annual Energy Outlook},'' 2019.
  [Online]. Available: \url{https://www.eia.gov/outlooks/aeo/pdf/aeo2019.pdf}
\BIBentrySTDinterwordspacing

\bibitem{dvorkin2015uncertainty}
Y.~Dvorkin, M.~Lubin, S.~Backhaus, and M.~Chertkov, ``Uncertainty sets for wind
  power generation,'' \emph{IEEE Transactions on Power Systems}, vol.~31,
  no.~4, pp. 3326--3327, 2015.

\bibitem{lubin2019chance}
M.~Lubin, Y.~Dvorkin, and L.~Roald, ``Chance constraints for improving the
  security of ac optimal power flow,'' \emph{IEEE Transactions on Power
  Systems}, vol.~34, no.~3, pp. 1908--1917, 2019.

\bibitem{roald2015security}
L.~Roald, F.~Oldewurtel, B.~Van~Parys, and G.~Andersson, ``Security constrained
  optimal power flow with distributionally robust chance constraints,''
  \emph{arXiv preprint arXiv:1508.06061}, 2015.

\bibitem{hodge2011wind}
B.-M. Hodge and M.~Milligan, ``Wind power forecasting error distributions over
  multiple timescales,'' in \emph{2011 IEEE power and energy society general
  meeting}.\hskip 1em plus 0.5em minus 0.4em\relax IEEE, 2011, pp. 1--8.

\end{thebibliography}

\appendix

\section{Appendix}
{
\subsection{Proxy of the social welfare function}\label{Appendix_proxy}
Given that means to achieve an RPS goal are chosen only after the goal itself is fixed and the goal is set at levels well above what would be the market outcome, optimal renewables build-out is achieved when the RPS goal is just met, i.e. the RES generation resulting from the RES investment does not exceed the RPS goal. 
This binding RPS constraint also means that feasible combinations of the generation types, and hence also  possible combinations of the tariff and RES incentives are limited. 
Furthermore, under RPS requirements, the unconstrained optimality conditions, such as the marginal social value of a generator equaling the social cost of adding it, do not hold: the capital costs of renewables investment substantially exceeds the social value of that generator. 
In other words, in the TL optimization proposed in this paper, the feasible region of the downstream (ML and LL) problems is confined by the constraints of the UL problem. When the problem of the state regulator is overly constrained by the given RPS goal, very few combinations of decisions are available to not only the decision makers in the UL problem, but also to the ones in the ML and LL problems. 
Consequently, minimizing total payments (tariff and incentive) is likely to approximate `social welfare maximization.' 

Therefore, when the elements of the social welfare function are not fully available (e.g. utility function of consumers due to privacy concerns, over-/under-estimated cost parameters of power utility), the `payment minimization' in \eqref{obj_payment} can be a fair proxy and adopted instead of the objective function in \eqref{Eq:RegObj}: 
\begin{align}    
    \begin{split}
        & \min_{\Xi^{\mathrm{R}}}  O^{\mathrm{R}}_s \coloneqq \hspace{-0.5mm} \sum_{t\in{\cal T}}\sum_{e\in{\cal E}} \omega_e \Big[ \sum_{n\in{\cal N}_s} d_{nte} \pi_{nt}  
        + \hspace{-4.5mm}\sum_{i\in\hat{\cal I}^{\mathrm R}\cup{\cal I}^{\mathrm R}} \hspace{-3mm}g_{ite} \tau^{\mathrm{e}}  \Big]
        \\&\hspace{20mm}
        + \hspace{-0.5mm}\sum_{i\in\hat{\cal I}^{\mathrm R}}  g^{\mathrm{max}}_i  \tau^{\mathrm{c}}
    \end{split}\label{obj_payment}
  \end{align}

\begin{table}[b]                      
    \small
    \centering
    \vspace{-3mm}
    	
        \setcounter{table}{0}
        \renewcommand{\thetable}{A\arabic{table}}
        \captionsetup{justification=centering, labelsep=period, font=footnotesize, textfont=sc}
        \caption{Comparison of regulatory decisions between they payment minimization and social welfare maximization objective functions \vspace{-0mm}}  
        \begin{tabular}{ P{1.5mm} P{3.5mm} | R{6mm} R{6mm} R{6mm} | R{6mm} R{6mm} R{6mm} | R{9.5mm} }
            \toprule
            &&\multicolumn{3}{c|}{Payment Min.*}&\multicolumn{3}{c|}{Social welfare Max.*}& \%change\\

            $\tau^{\mathrm{e}}$&$\tau^{\mathrm{c}}$&$\pi^{\mathrm{On}}$&$\pi^{\mathrm{Off}}$&$\mathbb{E}[\pi]$ &$\pi^{\mathrm{On}}$&$\pi^{\mathrm{Off}}$&$\mathbb{E}[\pi]$ &in $\mathbb{E}[\pi]$\\
            \midrule
            0&0&20.38&18.09&18.94&20.07&17.82&18.66&-1.49\%\\
            10&0&17.76&15.77&16.52&17.83&15.83&16.58&0.42\%\\
            20&0&15.22&13.52&14.16&15.22&13.52&14.15&0.01\%\\
            30&0&12.50&11.12&11.64&12.53&11.13&11.66&0.21\%\\
            40&0&9.63&8.57&8.969&9.62&8.55&8.95&-0.13\%\\
            50&0&7.01&6.23&6.527&7.09&6.31&6.60&1.26\%\\
            60&0&4.37&3.89&4.076&4.40&3.92&4.10&0.61\%\\
            0&100&19.60&17.39&18.22&19.59&17.39&18.22&0.00\%\\
            0&200&18.9&16.78&17.57&19.07&16.93&17.73&0.91\%\\
            0&300&18.02&16.01&16.76&18.26&16.21&16.98&1.34\%\\
            0&400&17.25&15.32&16.04&17.22&15.29&16.02&-0.16\%\\
            0&500&16.44&14.60&15.29&16.51&14.60&15.31&0.15\%\\
            0&600&15.67&13.92&14.58&15.64&13.90&14.55&-0.18\%\\
            \bottomrule
            \multicolumn{9}{l}{\scriptsize *Units: $\tau^{\mathrm{e}} \mathrm{[\$/MWh]}$, $\tau^{\mathrm{c}} \mathrm{[\$/kW]}$,
            $\pi^{\mathrm{On}}, \pi^{\mathrm{Off}} \mathrm{[\$/MWh]}$}
        \end{tabular}\label{Table_comp1}
    \vspace{-0mm}
\end{table}

To verify the validity of the proxy via `payment minimization', we carried out simulations of the TL optimization in Section \ref{sec:case_nh} where the original objective function in \eqref{Eq:RegObj} is replaced with \eqref{obj_payment}. Table~\ref{Table_comp1} compares the regulatory decisions between the two objectives. Both cases attained almost the same regulatory decision variables ($\tau^{\mathrm{e}},\tau^{\mathrm{c}},\pi_{nt}$), which differ by $1.5\%$ at most.

The similarity between `payment minimization' and `social welfare maximization' can also be found in the operating conditions and decisions of the power utility and the corresponding revenue and costs. For example, to meet the RPS goal set by the state regulator, the power utility installed 602 MW of wind generation resources in both cases, while further installation exceeding the RPS target is prohibited by its relatively expensive capital cost. Although the power utility's revenue is dependent on the market outcomes and generation costs and  may vary with the regulator's objective (payment min./social welfare max.) and RES incentives, it is notable that the total payment is set by the regulator such that the total revenue is equal to the exact amount to cover the investment and operation costs (see Table~\ref{Table_comp2}). In other words, regardless of the operation strategy chosen by the power utility, the state regulator selects an optimal combination of the consumer tariff and RES incentives so that the cost recovery is guaranteed, while  also achieving its RPS goal, i.e. $\mathrm{TP}\!+\!\mathrm{EP}\!\ge\!\mathrm{IC}\!+\!\mathrm{GC}$, where 
$\mathrm{TP}$ is a total payment ($\sum_{t\in{\cal T}}\sum_{n\in{\cal N}_s}\pi_{nt}d_{nte} + \sum_{t\in{\cal T}}\hspace{-0.5mm}\sum_{i\in\hat{\cal I}^{\mathrm R}\cup{\cal I}^{\mathrm R}}  \tau^{\mathrm{e}} g_{ite} + \hspace{-0.5mm}\sum_{i\in\hat{\cal I}^{\mathrm R}} \tau^{\mathrm{c}} g^{\mathrm{max}}_i$),
$\mathrm{EP}$ is a profit from the wholesale market ($
\sum_{t\in{\cal T}}  \hspace{-0.5mm}\sum_{n\in{\cal N}_s} \hspace{-0.5mm} \lambda_{nte} d_{nte}\hspace{-0mm} -  \hspace{-0mm}\sum_{t\in{\cal T}}\sum_{i \in {\cal I}\cup\hat{\cal I}} \hspace{-0.5mm}\lambda_{n(i),t,e} {g}_{ite}$),
$\mathrm{IC}$ is an investment cost ($\sum_{i\in\hat{\cal I}} C^{\mathrm{inv}}_i g^{\mathrm{max}}_{i}$), and 
$\mathrm{GC}$ is a generation cost ($\sum_{i\in{\cal I}}\sum_{t\in{\cal T}} C^{\mathrm{g}}_i g_{ite}$). 
Therefore, the cost recovery constraint in Eq.~(1d) is binding and the `payment minimization' and `social welfare maximization' are aligned as long as the RPS goal is externally set and enforced as a hard constraint. This provides a useful insight to the state regulator that `payment minimization' can be adopted as a proxy of `social welfare maximization' if the state regulator cannot fully identify each element of the welfare function.

\begin{table}[t]                      
    \small
    \centering

        \renewcommand{\thetable}{A\arabic{table}}
        \captionsetup{justification=centering, labelsep=period, font=footnotesize, textfont=sc}
        \caption{Comparison of the power utility revenues and costs under the payment minimization and social welfare maximization objective functions. \vspace{-0mm}}  
        \begin{tabular}{ P{1.5mm} P{3.5mm} | P{5.2mm} P{5.2mm} P{5.2mm} P{5.2mm} | P{5.2mm} P{5.2mm} P{5.2mm} P{5.2mm} }
            \toprule            
            &&\multicolumn{4}{c|}{Payment Min.*}&\multicolumn{4}{c}{Social welfare Max.*}\\
            $\tau^{\mathrm{e}}$&$\tau^{\mathrm{c}}$&TP&EP&IC&GC&TP&EP&IC&GC\\
            \midrule
            0&0&0.557&0.062&0.412&0.207&0.562&0.016&0.412&0.165\\
            10&0&0.560&0.048&0.412&0.196&0.562&0.081&0.412&0.230\\
            20&0&0.564&0.031&0.412&0.183&0.564&0.175&0.412&0.327\\
            30&0&0.564&0.050&0.412&0.201&0.565&0.097&0.412&0.250\\
            40&0&0.559&0.053&0.412&0.200&0.559&0.091&0.412&0.238\\
            50&0&0.561&0.044&0.412&0.193&0.564&0.078&0.412&0.229\\
            60&0&0.563&0.039&0.412&0.190&0.564&0.032&0.412&0.184\\
            0&100&0.557&0.059&0.412&0.204&0.557&0.041&0.412&0.186\\
            0&200&0.559&0.048&0.412&0.195&0.564&0.028&0.412&0.180\\
            0&300&0.557&0.060&0.412&0.204&0.563&0.051&0.412&0.202\\
            0&400&0.557&0.060&0.412&0.205&0.556&0.011&0.412&0.155\\
            0&500&0.556&0.071&0.412&0.215&0.557&0.035&0.412&0.179\\
            0&600&0.557&0.059&0.412&0.204&0.556&0.025&0.412&0.169\\
            \bottomrule                        
            \multicolumn{10}{l}{\scriptsize *TP: Total payment (tariff \& incentives), EP: Electricity market profit,}\\
            \multicolumn{10}{l}{\scriptsize \hspace{0.5mm}IC: Investment cost (prorated value), GC: Generation cost (in-state)}
            \\
            \multicolumn{10}{l}{\scriptsize *Unit: $\tau^{\mathrm{e}} \mathrm{[\$/MWh]}$, $\tau^{\mathrm{c}} \mathrm{[\$/kW]}$,
            $\mathrm{TP, EP, IC, GC~[\$M/day]}$.}
            \vspace{-4mm}
        \end{tabular}\label{Table_comp2}        
\end{table}
}

\subsection{Bi-level reformulation of Eq.~\eqref{Eq:bilevel} }
The complete bi-level reformulation of the proposed TL problem in Eq.~\eqref{Eq:bilevel} is as follows:

\begin{table*}[t]                      
    
    \centering
    	\renewcommand{\thetable}{A\arabic{table}}
    	\vspace{-3mm}
        \captionsetup{justification=centering, labelsep=period, font=footnotesize, textfont=sc}
        \captionof{table}{Inverse cumulative density (quantile) function of different probabilistic distributions for the second order conic reformulation of chance constraints.\vspace{-0mm}}
        \begin{tabular}{ L{27mm} | P{57mm} | P{14mm} P{22mm} | P{14mm} P{22mm} }    
            \toprule
            \multirow{2}{*}{Distribution}&\multirow{2}{*}{Quantile function (inverse CDF)}&\multirow{2}{*}{$\Phi^{-1}(95\%)$} &Allocated &\multirow{2}{*}{$\Phi^{-1}(97\%)$} &Allocated\\
            &&&Reserve [MW] && Reserve [MW]\\
            \midrule
            Standard Normal & $\Phi^{-1}(1-\eta) ={\sqrt {2}}\mathrm{erf} ^{-1}(1-2\eta ) $ & $1.645$ & $212.87$ & $1.881$ &$278.93$\\            
            $\text{Student's t (DOF$^\dagger$=4)}$ & 
            $\Phi^{-1}(1-\eta) = (1-2\eta) \sqrt{\frac{\cos{\frac{1}{3}\arccos{\sqrt{4\eta(1-\eta)}}}}{\sqrt{4\eta(1-\eta)}}}
            $ & $2.132$ & $274.35$ & $2.601$&$337.76$\\
            Logistic (scale = 1) & 
            $\Phi^{-1}(1-\eta) = \ln{\left(\frac{1-\eta}{\eta}\right)}
            $ & $2.944$ & $432.32$ & $3.476$&$444.50$\\
            Cauchy-Lorentz & 
            $\Phi^{-1}(1-\eta) = \tan{\left(\pi\left(\frac{1}{2}-\eta \right)\right)}
            $ & $6.314$ & $823.41$ & $10.579$ &$838.84$\\
            \bottomrule
            \multicolumn{3}{l}{\footnotesize{$^\dagger$DOF: degree of freedom}}
            \vspace{-3mm}
        \end{tabular}    
        \label{table:quantile}
\end{table*}

\begin{subequations}
\begin{align}
    \begin{split}
        & \max_{\Xi^{\mathrm{R}}} {O}^{\mathrm{R}}_s\! \coloneqq\! \sum_{e\in{\cal E}} \omega_e\Big[        
            \sum_{t\in{\cal T}}\sum_{n \in {\cal N}_s}\Big(\!M_{nte} d_{nte} \!-\! \frac{1}{2} N d_{nte}^2 \!-\! \pi_{nt} d_{nte} \Big)\hspace{-12mm}        
        \\& \hspace{5mm}        
            + \sum_{t\in{\cal T}}\sum_{n\in{\cal N}_s}\pi_{nt}d_{nte}  
            + \sum_{t\in{\cal T}}\hspace{-0.5mm}\sum_{i\in\hat{\cal I}^{\mathrm R}\cup{\cal I}^{\mathrm R}}  \hspace{-3mm}\tau^{\mathrm{e}} g_{ite}           
            + \hspace{-0.5mm}\sum_{i\in\hat{\cal I}^{\mathrm R}} \tau^{\mathrm{c}} g^{\mathrm{max}}_i           \hspace{-12mm}
        \\& \hspace{5mm}
        - 
            \hspace{-1mm}\sum_{t\in{\cal T}} \Big( \hspace{-0.5mm}\sum_{n\in{\cal N}_s} \hspace{-1.5mm} \lambda_{nte} d_{nte}\hspace{-1mm}
            -  \hspace{-2mm}\sum_{i \in {\cal I}\cup\hat{\cal I}} \hspace{-1.5mm}\lambda_{n(i),t,e} {g}_{ite} \!\Big)         
        \\& \hspace{5mm}        
        - 
            \sum_{i\in\hat{\cal I}} C^{\mathrm{inv}}_i g^{\mathrm{max}}_{i}
            - \sum_{i\in{\cal I}}\sum_{t\in{\cal T}} C^{\mathrm{g}}_i g_{ite}        
        \\&\hspace{5mm}         
        - 
            \sum_{t\in{\cal T}}\hspace{-0.5mm}\sum_{i\in\hat{\cal I}^{\mathrm R}\cup{\cal I}^{\mathrm R}}  \hspace{-3mm}\tau^{\mathrm{e}} g_{ite} 
            - \hspace{-0.5mm}\sum_{i\in\hat{\cal I}^{\mathrm R}} \tau^{\mathrm{c}} g^{\mathrm{max}}_i         
    \end{split}
\end{align}
\begin{align}
     & \sum_{t \in {\cal T}} \sum_{i\in{\cal I}_s^{\mathrm R}\cup\hat{\cal I}_s^{\mathrm R}} \!\!\!\!{g}_{ite} \tau^{\mathrm{e}} +   \sum_{i\in\hat{\cal I}_s^{\mathrm R}} {g}^{\mathrm{max}}_{i} \tau^{\mathrm{c}}  \le  B^{\mathrm{P}}_s,~\forall e\in{\cal E},\\
% \end{align}
% \begin{align}
     & \sum_{t \in {\cal T}} \sum_{i \in {\cal I}_s^{\mathrm R}\cup \hat{\cal I}_s^{\mathrm R}} \!\!{g}_{ite} \ge \kappa_s \sum_{t \in {\cal T}} \sum_{n\in{\cal N}_s} {d}_{nte},~\forall e\in{\cal E},\\
     & O^{\mathrm{U}}({g}^{\max}_i, {g}_{ite}) \ge 0,\\
     & \pi_{nt} = 
     \begin{cases}
        &\pi^{\mathrm{on}}_n,\quad\forall t \in {\cal T}^{\mathrm{on}},\\
        &\pi^{\mathrm{off}}_n,\quad\forall t \in {\cal T}^{\mathrm{off}},
    \end{cases}\\
    & \Xi^{\mathrm{U}} \cup \Xi^{\mathrm{W}} \in \arg\Big[ \nonumber\\
    \begin{split}             
        &\max_{\Xi^{\mathrm{U}}} O^{\mathrm{U}}\!\coloneqq\!
        \hspace{-0.0mm}\sum_{e\in{\cal E}} \omega_e \Big[\sum_{t\in{\cal T}}\sum_{n\in{\cal N}_s}\pi_{nt}d_{nte}  
        + \sum_{t\in{\cal T}}\hspace{-0.5mm}\sum_{i\in\hat{\cal I}^{\mathrm R}\cup{\cal I}^{\mathrm R}}  \hspace{-3mm}\tau^{\mathrm{e}} g_{ite}   \hspace{-5mm}
        \\&\hspace{0mm}
        + \hspace{-0.5mm}\sum_{i\in\hat{\cal I}^{\mathrm R}} \tau^{\mathrm{c}} g^{\mathrm{max}}_i  
        - \hspace{-1mm}\sum_{t\in{\cal T}} \Big( \hspace{-0.5mm}\sum_{n\in{\cal N}_s} \hspace{-1.5mm} \lambda_{nte} d_{nte}\hspace{-1mm}
        -  \hspace{-2mm}\sum_{i \in {\cal I}\cup\hat{\cal I}} \hspace{-1.5mm}\lambda_{n(i),t,e} {g}_{ite} \!\Big) \hspace{-5mm}
        \\&\hspace{0mm}
        - \sum_{i\in\hat{\cal I}} C^{\mathrm{inv}}_i g^{\mathrm{max}}_{i}
        - \sum_{i\in{\cal I}}\sum_{t\in{\cal T}} C^{\mathrm{g}}_i g_{ite}\Big]\hspace{-5mm}
    \end{split}\\
    \begin{split}
        & \overline{g}_{ite}
        -\alpha_{it} \Big( \hspace{-0.0mm} \sum_{j\in{\cal I}^{\mathrm{R}}_{s}} \hspace{-1.0mm} G^{\mathrm{max}}_j \upsilon_{jte} 
        \!+ \hspace{-1.0mm} \sum_{j\in\hat{\cal I}^{\mathrm{R}}_{s}} \hspace{-1.0mm} g^{\mathrm{max}}_j \upsilon_{jte}\Big)
        \\& \hspace{3.5mm}+ \Phi^{-1}(1\!-\!\eta){\Stdev[\bm{g}_{ite}]} 
        \le {G}_{i}^{\mathrm{max}},~\forall i \in  {\cal I}^{\mathrm{C}}_s, t \in {\cal T},\!\! 
    \end{split}\\
% \end{align}
% \begin{align}
    \begin{split}
        & \overline{g}_{ite}
        -\alpha_{it} \Big( \hspace{-0.0mm} \sum_{j\in{\cal I}^{\mathrm{R}}_{s}} \hspace{-1.0mm} G^{\mathrm{max}}_j \upsilon_{jte} 
        \!+ \hspace{-1.0mm} \sum_{j\in\hat{\cal I}^{\mathrm{R}}_{s}} \hspace{-1.0mm} g^{\mathrm{max}}_j \upsilon_{jte}\Big)
        \\& \hspace{3.5mm}+ \Phi^{-1}(1\!-\!\eta){\Stdev[\bm{g}_{ite}]} 
        \le {g}_{i}^{\mathrm{max}},~\forall i \in  \hat{\cal I}^{\mathrm{C}}_s, t \in {\cal T}, 
    \end{split}\\
    \begin{split}
        & \overline{g}_{ite}
        -\alpha_{it} \Big( \hspace{-0.0mm} \sum_{j\in{\cal I}^{\mathrm{R}}_{s}} \hspace{-1.0mm} G^{\mathrm{max}}_j \upsilon_{jte} 
        \!+ \hspace{-1.0mm} \sum_{j\in\hat{\cal I}^{\mathrm{R}}_{s}} \hspace{-1.0mm} g^{\mathrm{max}}_j \upsilon_{jte}\Big)
        \\& \hspace{3.5mm}- \Phi^{-1}(1\!-\!\eta){\Stdev[\bm{g}_{ite}]} 
        \ge {G}_{i}^{\mathrm{min}},~\forall i \in  {\cal I}^{\mathrm{C}}_s, t \in {\cal T}, \!
    \end{split}\\
    \begin{split}
        & \overline{g}_{ite}
        -\alpha_{it} \Big( \hspace{-0.0mm} \sum_{j\in{\cal I}^{\mathrm{R}}_{s}} \hspace{-1.0mm} G^{\mathrm{max}}_j \upsilon_{jte} 
        \!+ \hspace{-1.0mm} \sum_{j\in\hat{\cal I}^{\mathrm{R}}_{s}} \hspace{-1.0mm} g^{\mathrm{max}}_j \upsilon_{jte}\Big)
        \\& \hspace{3.5mm}- \Phi^{-1}(1\!-\!\eta){\Stdev[\bm{g}_{ite}]} 
        \ge \Gamma_i {g}_{i}^{\mathrm{max}},~\forall i \in  \hat{\cal I}^{\mathrm{C}}_s, t \in {\cal T}, \hspace{-10mm}
    \end{split}\\
    & {H}^{\mathrm{min}}_{i} \!\!\le\! \overline{g}_{ite}\!\!-\!\overline{g}_{i,t-1,e} \! \le \!{H}^{\mathrm{max}}_{i}\!\!,~\!\forall i \!\in \! {\cal I}^{\mathrm{C}}_s\!\cup\!\hat{\cal I}^{\mathrm{C}}_s, t \in \!{\cal T}\!, e\in\!{\cal E}\!,\hspace{-2mm}\\      
    & \sum_{i \in {\cal I}_s \cup {\hat{\cal{I}}_s}} \!\!\!\overline{g}_{ite} + \!{p}_{nte}^{\downarrow}  = \!d_{nte} ,\quad\forall t \in {\cal T},{n\in{\cal N}_s}, e\in{\cal E},\\
    & -\!{P}_n^{\downarrow,\mathrm{max}} \!\!\le {p}^{\downarrow}_{nte} \le {P}_n^{\downarrow,\mathrm{max}},~ \forall t \in {\cal T}\!, n\in{\cal N}_s, e\in{\cal E},\\
    &f_{lte} = \frac{1}{X_l} (\theta_{o(l),te}-\theta_{r(l),te}),~\forall {l \in {\cal L}},~t\in{\cal T},\!\!\\
    &\hspace{-1mm}\sum_{i \in {\cal I}_n\cup\hat{\cal I}_n } \hspace{-2.5mm}{g}_{ite}  + \hspace{-2mm}\sum_{l \vert r(l) = n } \hspace{-2.5mm} f_{lte} \!-\hspace{-3mm}\sum_{l \vert o(l) = n } \hspace{-2.5mm} f_{lte}\! = \!d_{nte}, \forall n \in {\cal N}, t\in{\cal T},\!\!\\
    &0 \le {g}_{ite} \le \overline{g}_{ite},~\forall i \in {\cal{I}}\cup\hat{\cal{I}},~t\in{\cal T}\!,\\
    &-{F}^{\mathrm{max}}_l \le f_{lte} \le {F}^{\mathrm{max}}_l,~\forall {l \in {\cal L}},~t\in{\cal T}\!,\!\!\\
% \end{align}
% \begin{align}
    & -\underline{\gamma}_{ite}\!+\!\overline{\gamma}_{ite}\!-\!\lambda_{n(i),te} \!=\! - C^{\mathrm g}_i ,~\forall i \in {\cal{I}},t \in {\cal T},\\
% \end{align}
% \begin{align}
    & \!-\underline{\delta}_{lte}+\overline{\delta}_{lte}+\xi_{lte} \!-\!\lambda_{r(l),t,e}+\lambda_{o(l),t,e} = 0 ,\forall l \in {\cal{L}},t \in {\cal T},
\end{align}
\begin{align}
    & -\!\!\!\!\sum_{l \vert o(l) = n}\! \frac{\xi_{lte}}{X_l}\!+\!\!\!\!\!\sum_{l \vert r(l) = n}\! \frac{\xi_{lte}}{X_l} = 0, ~\forall n \in {\cal{N}},t \in {\cal T},\\
    \begin{split}
        & \sum_{t\in{\cal T}} \sum_{i \in {\cal I}\cup\hat{\cal I}} - C_i^{\mathrm{g}} {g}_{ite} =\sum_{t\in {\cal T}} \Big( \sum_{i \in {\cal I}\cup\hat{\cal I}} \! \overline{g}_{ite} \overline{\gamma}_{ite}
        \\
        &\hspace{6mm}- \!\!\sum_{n \in {\cal N}} \!D_{nte}^{\mathrm{p}} \lambda_{nte}   + \sum_{l \in {\cal L}} \! \big({F}^{\mathrm{max}}_l \overline{\delta}_{lte}\!+\!{F}^{\mathrm{max}}_l \underline{\delta}_{lte} \big) \Big),       
    \end{split}\\
    &\quad\forall e\in{\cal E}\hspace{1.3mm}
        \hspace{3mm}\Big].\nonumber
\end{align}\end{subequations}

\vspace{-3mm}
{
\subsection{Extension for Non-Gaussian probabilistic distributions}\label{Appendix_NonGuassian}

While the Gaussian assumption makes it possible to accurately capture first- and second-order moments (e.g. mean and standard deviation) of the RES forecast errors (see the reports in \cite{dvorkin2015uncertainty,lubin2019chance}), the accuracy of probabilistic assumptions can be further improved by adopting probabilistic distributions such as Student's t, Logistic, and Cauchy-Lorentz distributions, which are relevant for modeling RES uncertainty \cite{roald2015security,hodge2011wind}. The proposed TL can be modified to handle non-Gaussian distributions by replacing the quantile function, $\Phi^{-1}(1-\eta)$, in Eqs.~\eqref{Eq:gen_UB_Conic}\textendash\eqref{Eq:gen_LB_New_Conic}. Fig.~\ref{fig:quantile} shows quantile functions of different distributions as a function of security parameter $\eta$ and Table~\ref{table:quantile} lists their analytical expression and the corresponding allocated security margin (reserve) computed based on Eq.~\eqref{Eq:CC_reform} as {\footnotesize$\sum_{i\in{\cal I}^{\mathrm{C}}\cup\hat{\cal I}^{\mathrm{C}}}\alpha_{it}\Phi^{-1}(1\!-\!\eta)\!\sqrt{\sum_{j\in{\cal{I}}_{s(i)}^{\mathrm{R}}}({{G}^{\mathrm{max}}_{j}} \sigma_{jte})^2 \!+\!\! \sum_{j\in\hat{\cal{I}}_{s(i)}^{\mathrm{R}}}({{g}^{\mathrm{max}}_{j}} \sigma_{jte})^2 }$}. Cauchy-Lorentz distribution has the greatest quantile value (most conservative) among the compared distributions, while the standard normal distribution has the lowest value (least conservative). As a result, greater margins  have been allocated for distributions with a higher quantile value. Additionally, the case with $\eta\!=\!3\%$ yields greater security margins than the case with relatively less conservative  tolerance ($\eta\!=\!5\%$).

\begin{figure}[h]
    \setcounter{figure}{0}
    \renewcommand{\thefigure}{A\arabic{figure}}
    \centering
    \vspace{-0mm}
    \includegraphics[width=0.95\columnwidth]{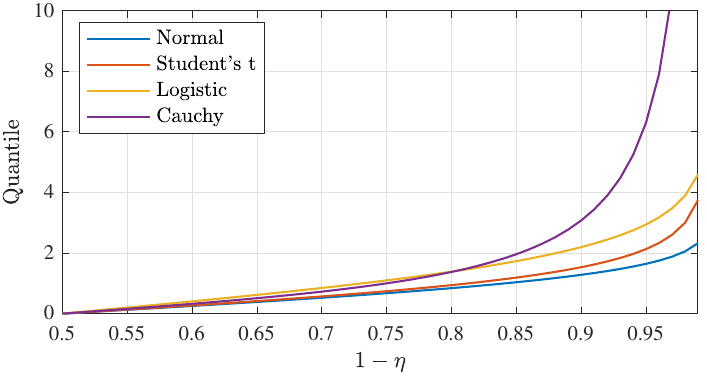}
    \vspace{-1mm}
    \captionof{figure}{\small Value of the inverse cumulative density (quantile) function for different probabilistic distributions  and different values of  $\eta$.}
    \label{fig:quantile}
\end{figure} 
}
\end{document}